\documentclass[iop,revtex4]{emulateapj}
\slugcomment{Accepted to AJ: 23 March 2017}

\usepackage[backref,breaklinks,colorlinks,citecolor=blue]{hyperref}
\usepackage[all]{hypcap}
\usepackage{graphicx} 
\usepackage{mathrsfs} 
\usepackage{natbib}
\usepackage{amsmath}
\usepackage{float}
\usepackage{amssymb}
\usepackage{color}
\usepackage{captcont}

\begin{document}

\title{An ALMA Survey of Protoplanetary Disks in the $\sigma$ Orionis Cluster}

\author{M. Ansdell$^{1}$, J. P. Williams$^{1}$, C.F. Manara$^{2}$, A. Miotello$^{3}$, S. Facchini$^{4}$, N. van der Marel$^{1}$, L. Testi$^{5,6}$, E. F. van Dishoeck$^{3,4}$}
 
\affil{$^1$Institute for Astronomy, University of Hawai`i at M\={a}noa, Honolulu, HI, USA}
\affil{$^2$Scientific Support Office, Directorate of Science, European Space Research and Technology Centre (ESA/ESTEC), Keplerlaan 1, 2201 AZ Noordwijk, The Netherlands}
\affil{$^3$Leiden Observatory, Leiden University, PO Box 9513, 2300 RA Leiden, The Netherlands}
\affil{$^4$Max-Plank-Institut f\"{u}r Extraterrestrische Physik, Giessenbachstra\ss e 1, D-85748 Garching, Germany}
\affil{$^5$INAF-Osservatorio Astrofisico di Arcetri, Largo E. Fermi 5, I-50125 Firenze, Italy}
\affil{$^6$European Southern Observatory, Karl-Schwarzschild-Str. 2, D-85748 Garching bei M\"{u}nchen, Germany}


\begin{abstract}

The $\sigma$~Orionis cluster is important for studying protoplanetary disk evolution, as its intermediate age ($\sim$$3$--5~Myr) is comparable to the median disk lifetime. We use ALMA to conduct a high-sensitivity survey of dust and gas in 92 protoplanetary disks around $\sigma$~Orionis members with $M_{\ast}\gtrsim0.1~M_{\odot}$. Our observations cover the 1.33~mm continuum and several CO $J=2$--1 lines: out of 92 sources, we detect 37 in the mm continuum and six in $^{12}$CO, three in $^{13}$CO, and none in C$^{18}$O. Using the continuum emission to estimate dust mass, we find only 11 disks with $M_{\rm dust}\gtrsim10~M_{\oplus}$, indicating that after only a few Myr of evolution most disks lack sufficient dust to form giant planet cores. Stacking the individually undetected continuum sources limits their average dust mass to 5$\times$ lower than that of the faintest detected disk, supporting theoretical models that indicate rapid dissipation once disk clearing begins. Comparing the protoplanetary disk population in $\sigma$~Orionis to those of other star-forming regions supports the steady decline in average dust mass and the steepening of the $M_{\rm dust}$--$M_{\ast}$ relation with age; studying these evolutionary trends can inform the relative importance of different disk processes during key eras of planet formation. External photoevaporation from the central O9 star is influencing disk evolution throughout the region: dust masses clearly decline with decreasing separation from the photoionizing source, and the handful of CO detections exist at projected separations $>1.5$~pc. Collectively, our findings indicate that giant planet formation is inherently rare and/or well underway by a few Myr of age. 

\end{abstract}

\maketitle


\section{INTRODUCTION}
\label{sec-intro}

Planets are thought to form in so-called ``protoplanetary" disks around young stars within $\sim$5--10~Myr \citep{2011ARA&A..49...67W}. The resulting exoplanet population is diverse, as revealed by the {\it Kepler} transit survey \citep{2010Sci...327..977B} and long-term radial velocity (RV) surveys \citep{2010Sci...330..653H,2011arXiv1109.2497M}. However, certain trends are emerging; for example, intermediate-mass planets  (i.e., ``super-Earths" with masses between that of Earth and Neptune) appear to be an order of magnitude more abundant than gas giants (i.e., planets with masses several times that of Jupiter), at least for short orbital periods \citep{2012ApJS..201...15H,2013ApJ...766...81F,2013ApJ...770...69P,2016MNRAS.457.2877G}. To identify the origins of these trends, and thus better understand planet formation, we must survey the preceding protoplanetary disks. Indeed, exoplanet population synthesis models indicate that planetary properties and architectures are dictated by the initial dust and gas content of protoplanetary disks and their subsequent evolution \cite[e.g.,][]{2008Sci...321..814T,2012A&A...541A..97M, 2015A&A...575A..28B,2016ApJ...832...41M}. 

Sub-mm and mm wavelength surveys are particularly useful for probing the bulk dust and gas content of protoplanetary disks, as disk emission at these longer wavelengths can be optically thin. The first (sub-)mm surveys of star-forming regions made the initial steps in identifying trends in protoplanetary disk evolution that could potentially explain correlations seen in the exoplanet population \citep{2009ApJ...700.1502A,2011ApJ...736..135L,2013ApJ...771..129A,2013MNRAS.435.1671W,2015ApJ...806..221A}. Most notably, early surveys of Taurus disks \citep{2000prpl.conf..559N,2013ApJ...771..129A} revealed a positive dependence between disk dust mass ($M_{\rm dust}$) and host star mass ($M_{\ast}$), which could fundamentally explain the positive correlation between giant planet frequency and stellar mass \citep{2006ApJ...649..436E,2007ApJ...670..833J,2010ApJ...709..396B,2013A&A...549A.109B}. However, these initial disk surveys were often incomplete and limited by dust mass sensitivities of a few Earth masses. These constraints meant that it remained unclear whether (sub-)mm continuum emission systematically declines with age, reflecting steady disk dispersal and/or grain growth in protoplanetary disks \cite[e.g.,][]{2012M&PS...47.1915W}. Moreover, none of these initial surveys probed bulk gas mass, as contemporary facilities lacked the sensitivity to detect faint line emission. 

Measuring both dust and gas content independently is essential for studying planet formation, as growing dust grains decouple from the gas and evolve differently, yet both components determine what types of planets may form in a disk. However, due to the challenges associated with estimating disk gas masses, the canonical interstellar medium (ISM) gas-to-dust ratio of $\sim$100 \citep{1978ApJ...224..132B} is often used to infer gas mass from dust mass, requiring an extrapolation of two orders of magnitude. Recent observations suggest that the inherited ISM ratio may actually decrease by an order of magnitude after just a few Myr of evolution \citep{2014ApJ...788...59W, 2016ApJ...828...46A}, although these calculations may be underestimated due to carbon depletion \cite[e.g.,][]{2016arXiv161201538M}. If gas is being depleted in disks (e.g., due to winds) then this may help to explain the lack of gas giants and prevalence of super-Earths seen in the exoplanet population \cite[e.g.,][]{2013ApJ...766...81F,2016MNRAS.457.2877G}. In this scenario, super-Earths would result when giant planet cores form in gas-depleted disks, prohibiting the cores from rapidly accreting gaseous envelopes \cite[e.g.,][]{2016ApJ...817...90L} as predicted by core accretion theory \cite[e.g.,][]{1996Icar..124...62P,2004ApJ...604..388I}. 

The enhanced sensitivity of the Atacama Large Millimeter/sub-millimeter Array (ALMA) now allows for efficient surveys of both dust and gas for large samples of protoplanetary disks across star-forming regions spanning the expected disk lifetime ($\sim$1--10~Myr). The first large-scale protoplanetary disk surveys conducted with ALMA include: \cite{2016ApJ...828...46A}, who carried out a near-complete survey of 89 protoplanetary disks in the young Lupus star-forming region \cite[$\sim$1--3~Myr at $\sim$150~pc;][]{2008hsf2.book..295C,2014A&A...561A...2A} with continuum and line sensitivities corresponding to $M_{\rm dust}$$\sim$$0.3$~$M_{\oplus}$ and $M_{\rm gas}$$\sim$$1.0$~$M_{\rm Jup}$, respectively; \cite{2016ApJ...827..142B}, who observed 106 disks in the more evolved Upper Sco region \cite[$\sim$5--10~Myr at 145~pc;][]{2002AJ....124..404P, 2012ApJ...746..154P} with sensitivities of $M_{\rm dust}$$\sim$$0.1$~$M_{\oplus}$; and \cite{2016ApJ...831..125P}, who surveyed 93 protoplanetary disks in the young Chamaeleon I region \cite[$\sim$2--3~Myr at 160~pc;][]{2008hsf2.book..169L} with sensitivities of $M_{\rm dust}$$\sim$0.2--0.8~$M_{\oplus}$.\footnote{These dust mass sensitivities correspond to $3\times$ the rms and use the assumptions described in Section~\ref{sec-dust} to convert (sub-)mm flux to dust mass.}

These ALMA disk surveys are beginning to reveal trends in protoplanetary disk evolution that can help to constrain planet formation theory and explain correlations seen in the exoplanet population. One of their clearest findings is that average disk dust mass does indeed decrease with age. \cite{2016ApJ...828...46A} showed that Lupus disks have a mean dust mass $\sim$3$\times$ higher than that of the older Upper Sco region, but are statistically indistinguishable from disks in the similarly aged Taurus region. Equivalently, \cite{2016ApJ...827..142B} found that the average $M_{\rm dust}/M_{\ast}$ ratio in Upper Sco is $\sim$4.5$\times$ lower than in Taurus. Interestingly, even in the younger regions, most disks lack sufficient dust to form the solid cores needed to build gas giants \cite[e.g., only 26\% of protoplanetary disks in Lupus have $M_{\rm dust}\gtrsim10~M_{\oplus}$;][]{2016ApJ...828...46A}. These findings point to significant global disk evolution during the first few Myr and imply that giant planet formation occurs rapidly and/or is rare.

Moreover, these ALMA disk surveys have confirmed the aforementioned $M_{\rm dust}$--$M_{\ast}$ relation initially seen in the pre-ALMA surveys of Taurus, while also revealing a steepening of the relation with age \citep{2016ApJ...828...46A,2016ApJ...831..125P}. This steepening would indicate that dust evolution occurs more rapidly around lower-mass stars, and can be compared to theoretical models to constrain the relative importance of different disk evolution processes \citep{2016ApJ...831..125P}. Finally, \cite{2016A&A...591L...3M} have combined estimates of disk mass ($M_{\rm disk}$) from ALMA with spectroscopic measurements of stellar mass accretion rates ($\dot{M}_{\rm acc}$) from VLT/X-Shooter in the Lupus star-forming region to provide the first observational confirmation of the $M_{\rm disk}$--$\dot{M}_{\rm acc}$ relation predicted by viscous disk evolution theory.

The $\sigma$~Orionis region \citep{2008hsf1.book..732W} is a particularly important target for studying disk evolution due to its intermediate age of $\sim$3--5~Myr \citep{2002A&A...382L..22O,2004MNRAS.347.1327O}, which is comparable to the median disk lifetime \citep{2011ARA&A..49...67W}. Only one-third of cluster members (92 sources) exhibit strong infrared (IR) excess indicative of a protoplanetary disk \citep{2007ApJ...662.1067H}. \cite{2013MNRAS.435.1671W} surveyed the region at 850~$\mu$m with JCMT/SCUBA-2, detecting just 8 disks at $\sim$$4~M_{\oplus}$ sensitivity; they also noted the remarkable diversity in the spectral energy distributions (SEDs) of their detections (see their Figure~3), indicating substantial and ongoing disk evolution. By stacking their individually non-detected disks, \cite{2013MNRAS.435.1671W} found a mean signal of 1.3~mJy at 4$\sigma$ significance, motivating more sensitive follow-up observations of this region.

We therefore use ALMA to conduct a high-sensitivity mm wavelength survey of all known protoplanetary disks in $\sigma$~Orionis in both dust and gas. We describe the sample in Section~\ref{sec-sample} and our ALMA observations in Section~\ref{sec-observations}. The continuum and line measurements are presented in Section~\ref{sec-results}, then converted to dust and gas masses in Section~\ref{sec-analysis}. We interpret our findings within the context of disk evolution in Section~\ref{sec-discussion} by identifying correlations with stellar and cluster properties as well as comparing our results to those found in other star-forming regions. This work is summarized in Section~\ref{sec-summary}.


\capstartfalse
\begin{figure}
\begin{centering}
\includegraphics[width=8.7cm]{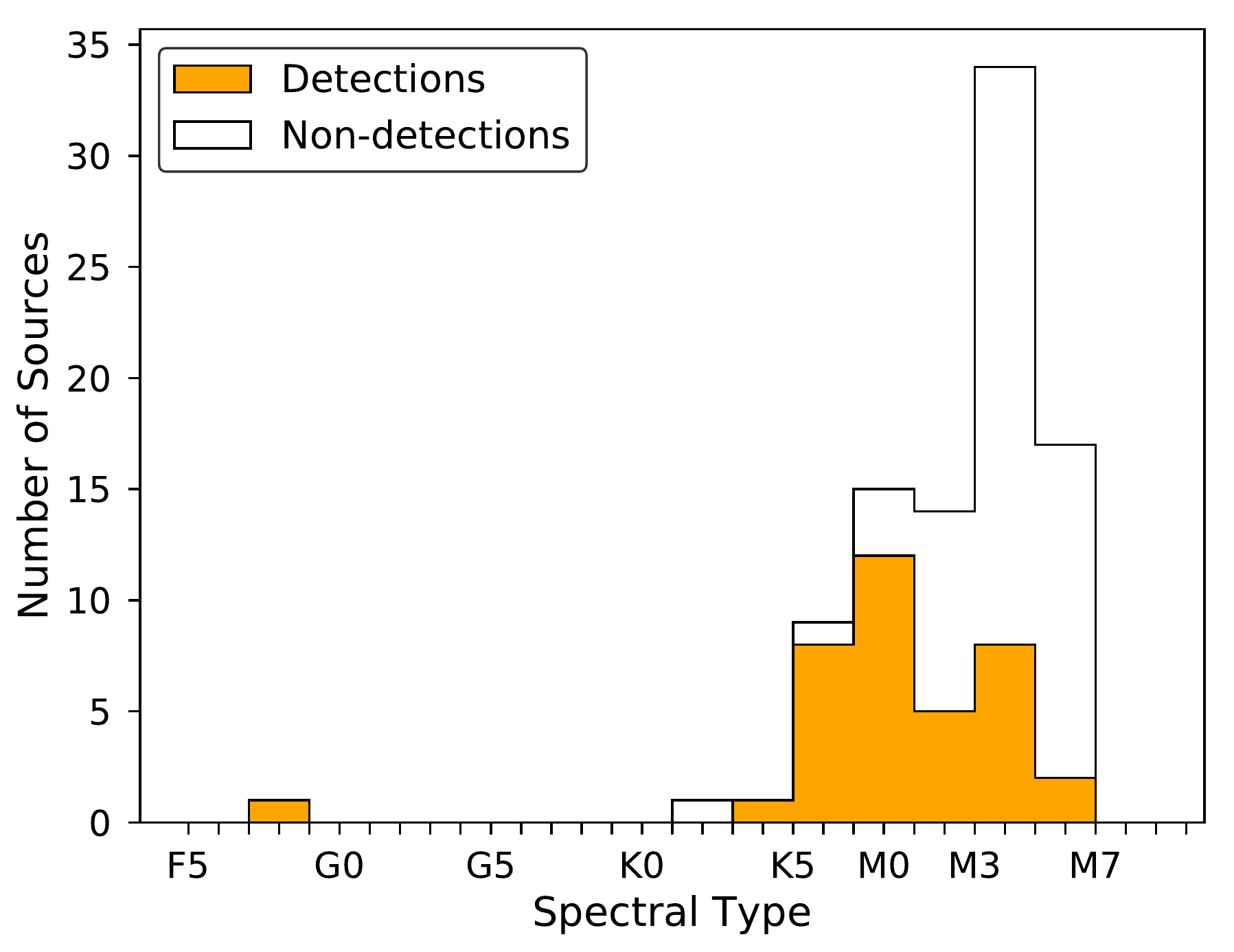}
\caption{\small Distribution of stellar spectral types for the sources in $\sigma$~Orionis targeted by our ALMA survey (Table~\ref{tab-cont}). The orange histogram shows continuum detections, while the open histogram shows continuum non-detections (Section~\ref{sec-cont}).}
\label{fig-spt}
\end{centering}
\end{figure}
\capstartfalse

\capstartfalse 
\begin{deluxetable*}{lllcrrrcr} 
\tabletypesize{\footnotesize} 
\centering 
\tablewidth{500pt} 
\tablecaption{Continuum Properties \label{tab-cont}} 
\tablecolumns{9}  
\tablehead{ 
 \colhead{Source} 
&\colhead{RA$_{\rm J2000}$} 
&\colhead{Dec$_{\rm J2000}$} 
&\colhead{SpT} 
&\colhead{Ref} 
&\colhead{$M_{\ast}$} 
&\colhead{$F_{\rm 1.33mm}$} 
&\colhead{rms} 
&\colhead{$M_{\rm dust}$} \\ 
 \colhead{} 
&\colhead{} 
&\colhead{} 
&\colhead{} 
&\colhead{} 
&\colhead{($M_{\odot}$)} 
&\colhead{(mJy)} 
&\colhead{(mJy beam$^{-1}$)} 
&\colhead{($M_{\oplus}$)} 
} 
\startdata 
1036 & 05:39:25.206 & -02:38:22.09 & K7.5 $\pm$ 0.5 & H14 & 0.67 $\pm$ 0.06 & 5.94 $\pm$ 0.15 & 0.16 & 26.45 $\pm$ 0.66 \\
1050 & 05:39:26.330 & -02:28:37.70 & M3.9 $\pm$ 2.0 & VJ & 0.19 $\pm$ 0.12 & -0.15 $\pm$ 0.15 & 0.15 & -0.67 $\pm$ 0.66 \\
1075 & 05:39:29.350 & -02:27:21.02 & M0.0 $\pm$ 1.5 & H14 & 0.62 $\pm$ 0.14 & 1.48 $\pm$ 0.15 & 0.16 & 6.57 $\pm$ 0.65 \\
1152 & 05:39:39.377 & -02:17:04.50 & M0.0 $\pm$ 1.0 & H14 & 0.62 $\pm$ 0.16 & 8.57 $\pm$ 0.17 & 0.18 & 38.16 $\pm$ 0.77 \\
1153 & 05:39:39.828 & -02:31:21.89 & K5.5 $\pm$ 1.0 & H14 & 0.91 $\pm$ 0.12 & 13.62 $\pm$ 0.16 & 0.18 & 60.66 $\pm$ 0.72 \\
1154 & 05:39:39.833 & -02:33:16.08 & M3.8 $\pm$ 2.0 & VJ & 0.25 $\pm$ 0.14 & 1.44 $\pm$ 0.15 & 0.16 & 6.43 $\pm$ 0.65 \\
1155 & 05:39:39.900 & -02:43:09.00 & K1.0 $\pm$ 2.5 & H14 & 1.71 $\pm$ 0.25 & -0.12 $\pm$ 0.15 & 0.15 & -0.54 $\pm$ 0.65 \\
1156 & 05:39:40.171 & -02:20:48.04 & K5.0 $\pm$ 1.0 & H14 & 0.96 $\pm$ 0.16 & 5.66 $\pm$ 0.15 & 0.18 & 25.21 $\pm$ 0.68 \\
1182 & 05:39:43.190 & -02:32:43.30 & M4.0 $\pm$ 2.0 & VJ & 0.20 $\pm$ 0.13 & 0.18 $\pm$ 0.15 & 0.15 & 0.82 $\pm$ 0.66 \\
1193 & 05:39:44.510 & -02:24:43.20 & M5.0 $\pm$ 2.0 & VJ & 0.10 $\pm$ 0.10 & -0.02 $\pm$ 0.15 & 0.15 & -0.09 $\pm$ 0.66
\enddata 
\tablenotetext{}{References: H14 = \cite{2014ApJ...794...36H}, R12 = \cite{2012A&A...548A..56R}, VJ = derived from $V-J$ colors (see Section~\ref{sec-sample}).} 
\tablenotetext{}{(This table is available online in its entirety in machine-readable form.)} 
\end{deluxetable*} 
\capstartfalse

\section{SAMPLE }
\label{sec-sample}

The $\sigma$~Orionis cluster consists of several hundred young stellar objects (YSOs) ranging from brown dwarfs to OB-type stars \cite[see review in][]{2008hsf1.book..732W}. The cluster is named after its brightest member, $\sigma$~Ori, a trapezium-like system whose most massive component is an O9 star. Cluster membership is defined by the Mayrit catalog \citep{2008A&A...478..667C}, which identifies 241 stars and brown dwarfs that are located within 30~arcmin of the $\sigma$~Ori system and exhibit known features of youth (X-ray emission, Li 6708\AA{} absorption, etc.). We adopt a cluster distance of 385~pc based on recent orbital parallax measurements of the $\sigma$~Ori triple system \citep{2016AJ....152..213S}. The low reddening towards this cluster, estimated at $E(B-V)\sim0.05$ \cite[e.g.][]{2008AJ....135.1616S}, makes it a valuable site for analyzing the evolution of young stars. 

Our sample consists of the 92 YSOs in $\sigma$~Orionis with IR excesses consistent with the presence of a protoplanetary disk. These sources are identified by cross-matching the Class~II and transition disk (TD) candidates from the {\it Spitzer} survey of \cite{2007ApJ...662.1067H} with the aforementioned Mayrit catalog \citep{2008A&A...478..667C}. Both catalogs are expected to be complete down to the brown dwarf limit. Disk classifications are based on the {\it Spitzer}/IRAC SED slope, as described in \cite{2007ApJ...662.1067H}. We also include in our sample a Class~I disk (1153), as it is located near the {\it Spitzer}/IRAC color cutoff for Class~II disks. 

The sources in our sample are presented in Table~\ref{tab-cont} with their stellar spectral types (SpT) and stellar masses ($M_{\ast}$). Spectral types were primarily taken from the homogenous sample of low-resolution optical spectra analyzed in \cite{2014ApJ...794...36H}, but supplemented with those from medium-resolution VLT/X-Shooter spectra when available from \cite{2012A&A...548A..56R}. For the 23 sources that lack spectroscopic information, we estimate their spectral types using an empirical relation between $V-J$ color and stellar spectral type; the relation was derived by measuring synthetic photometry from flux-calibrated VLT/X-Shooter spectra of YSOs with spectral types from G5 to M9.5, then performing a non-parametric fit of the $V-J$ color versus spectral type relation (Manara et al. 2017, submitted). For these sources with photometrically derived spectral types, we cautiously assume uncertainties of $\pm$ two spectral subtypes. Figure~\ref{fig-spt} shows the stellar spectral type distribution of our sample.

We estimate $M_{\ast}$ values for our sample by comparing their positions on the Hertzsprung-Russel (HR) diagram to the evolutionary models of \cite{2000A&A...358..593S}. In order to place our targets on the HR diagram, we convert their spectral types to stellar effective temperatures ($T_{\rm eff}$) and derive their stellar luminosities ($L_{\ast}$) from $J$-band magnitudes using the relations in \cite{2015ApJ...808...23H}. The uncertainties on $L_{\ast}$ are obtained by propagating the uncertainties on spectral type and bolometric correction, and thus on distance and optical extinction ($A_V$). We then calculate the uncertainties on $M_{\ast}$ using a Monte Carlo (MC) method, where we take the standard deviation of 1000 estimates of $M_{\ast}$, each calculated after randomly perturbing the derived values of $T_{\rm eff}$ and $L_{\ast}$ by their uncertainties.

\capstartfalse
\begin{figure*}
\begin{centering}
\includegraphics[width=18cm]{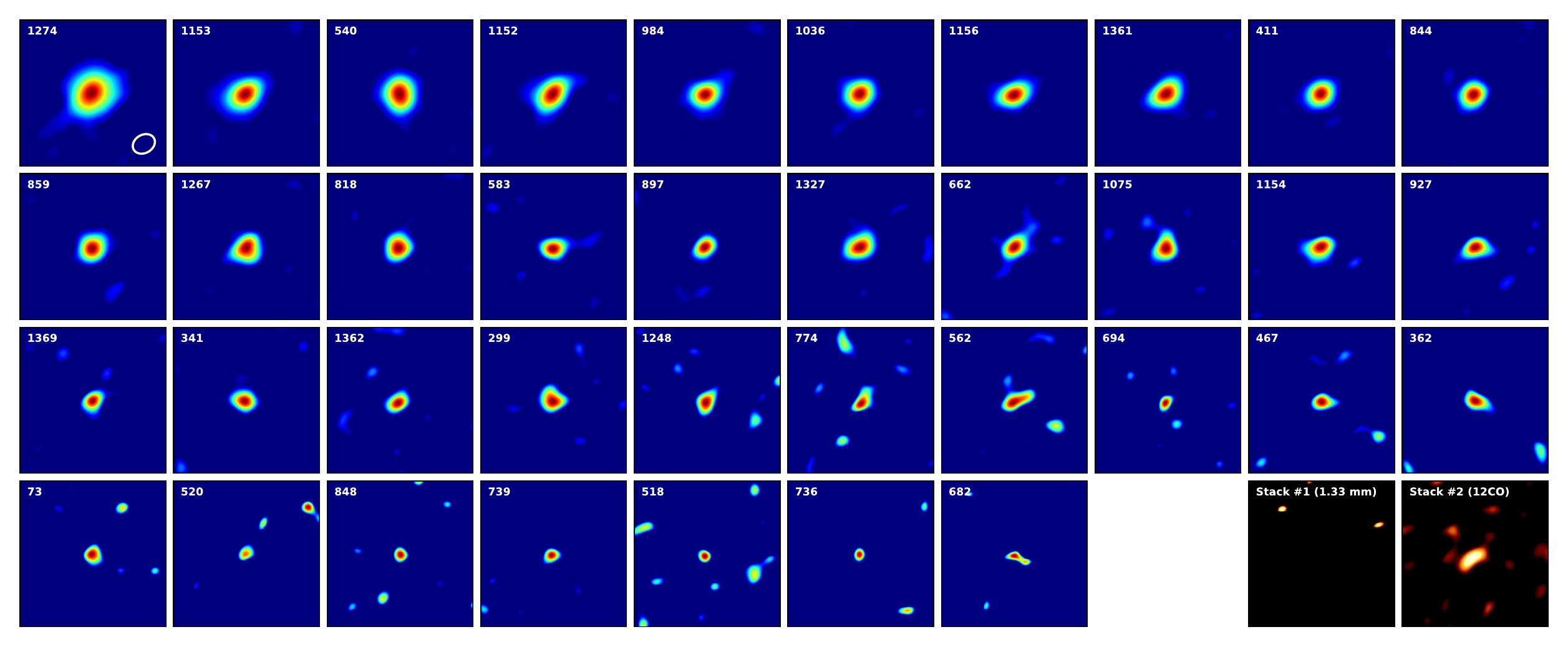}
\caption{\small Continuum images at 1.33~mm of the 37 detected disks in our $\sigma$~Orionis ALMA sample, ordered by decreasing flux density (as reported in Table~\ref{tab-cont}). The last two images show the stacked non-detections described in Section~\ref{sec-stacks}. Images are 2\arcsec$\times$2\arcsec and the typical beam size of 0\farcs31$\times$0\farcs25 (Section~\ref{sec-observations}) is shown in the first panel.}
\label{fig-cont}
\end{centering}
\end{figure*}
\capstartfalse


\section{ALMA OBSERVATIONS}
 \label{sec-observations}
 
Our Band 6 ALMA observations were obtained on 2016 July 30 and 31 during Cycle~3 (Project ID: 2015.1.00089.S; PI: Williams). The array configuration used 36 and 37 12~m antennas on July 30 and 31, respectively, with baselines of 15--1124~m on both runs. The correlator setup included two broadband continuum windows centered on 234.293 and 216.484~GHz with bandwidths of 2.000 and 1.875~GHz and channel widths of 15.625 and 0.976~MHz, respectively. The bandwidth-weighted mean continuum frequency was 225.676~GHz (1.33~mm). The spectral windows covered the $^{12}$CO (230.538~GHz), $^{13}$CO (220.399~GHz), and C$^{18}$O (219.560~GHz) $J=2$--1 transitions at velocity resolutions of 0.16-0.17~km~s$^{-1}$. These spectral windows were centered on 230.531, 220.392, and 219.554~GHz with bandwidths of 11.719~MHz and channel widths of 0.122~MHz.

On-source integration times were 1.2~min per object for an average continuum rms of 0.15~mJy~beam$^{-1}$ (Table~\ref{tab-cont}). This sensitivity was based on the JCMT/SCUBA-2 survey of $\sigma$~Orionis disks by \cite{2013MNRAS.435.1671W}, who found that stacking their individual non-detections revealed a mean 850~$\mu$m continuum signal of 1.3~mJy at 4$\sigma$ significance. The sensitivity of our ALMA survey was therefore chosen to provide $\sim$3--4$\sigma$ detections of such disks at 1.3~mm, based on an extrapolation of the 850~$\mu$m mean signal using a spectral slope of $\alpha=2$--3.

The raw data were pipeline calibrated by NRAO staff using the \texttt{CASA} package (version 4.5.3). The pipeline calibration included: absolute flux calibration with observations of J0522-3627 or J0423-0120; bandpass calibration with observations of J0510+1800 or J0522-3627; and gain calibration with observations of J0532-0307. We estimate an absolute flux calibration error of $\sim$10\% based on the amplitude variations of gain calibrators over time. 

We extract the continuum images from the calibrated visibilities by averaging over the continuum channels and cleaning with a Briggs robust weighting parameter of +0.5 for an average beam size of 0\farcs31$\times$0\farcs25 ($\sim$120$\times$95~AU at 385~pc). We extract $^{12}$CO, $^{13}$CO, and C$^{18}$O $J=2$--1 line channel maps from the calibrated visibilities by subtracting the continuum from the spectral windows containing line emission using the {\it uvcontsub} routine in \texttt{CASA}. Sources showing clear line emission were cleaned with a Briggs robust weighting parameter of +0.5. We find average rms values of 13, 14, and 11~mJy~beam$^{-1}$ within 1~km~s$^{-1}$ velocity channels for the $^{12}$CO, $^{13}$CO, and C$^{18}$O lines, respectively.


\section{ALMA RESULTS}
\label{sec-results}

\subsection{1.33~mm Continuum Emission  \label{sec-cont}}

Nearly all sources in our sample are unresolved, thus we measure continuum flux densities by fitting point-source models to the visibility data using the {\it uvmodelfit} routine in \texttt{CASA}. The point-source model has three free parameters: integrated flux density ($F_{\lambda}$), right ascension offset from the phase center ($\Delta\alpha$), and declination offset from the phase center ($\Delta\delta$). For the five resolved sources (1036, 1152, 1153, 1274, 540), we fit an elliptical Gaussian model instead, which has three additional free parameters: FWHM along the major axis ($a$), aspect ratio of the axes ($r$), and position angle (PA). We scale the uncertainties on the fitted parameters by the square root of the reduced $\chi^{2}$ value of the fit.

Table~\ref{tab-cont} presents the 1.33~mm continuum flux densities and associated uncertainties ($F_{\rm 1.33mm}$), where the uncertainties are statistical errors and do not include the 10\% flux calibration error (Section~\ref{sec-observations}). We detect only 37 out of the 92 observed sources at $>3\sigma$ significance (Figure~\ref{fig-cont}). For detections, the source locations in Table~\ref{tab-cont} are the fitted source centers output by {\it uvmodelfit}, while for non-detections they are simply the phase centers of the ALMA observations, which were chosen based on 2MASS positions. The average offsets from the phase centers for the detections are $\Delta\alpha=0\farcs057$ and $\Delta\delta=-0\farcs096$ (1.9 and $ -3.2$ pixels), both much smaller than the average beam size (Section~\ref{sec-observations}). We also note that only 5 out of the 37 continuum detections have photometrically derived spectral types, which are less precise than the spectroscopically determined spectral types (Section~\ref{sec-sample}).

\capstartfalse 
\begin{deluxetable*}{lrrrrrr} 
\tabletypesize{\footnotesize} 
\centering 
\tablewidth{500pt} 
\tablecaption{Gas Properties \label{tab-gas}} 
\tablecolumns{8}  
\tablehead{ 
 \colhead{Source} 
&\colhead{$F_{\rm 12CO}$} 
&\colhead{$F_{\rm 13CO}$} 
&\colhead{$F_{\rm C18O}$} 
&\colhead{$M_{\rm gas}$} 
&\colhead{$M_{\rm gas,min}$} 
&\colhead{$M_{\rm gas,max}$} \\ 
 \colhead{} 
&\colhead{(mJy~km~s$^{-1}$)} 
&\colhead{(mJy~km~s$^{-1}$)} 
&\colhead{(mJy~km~s$^{-1}$)} 
&\colhead{($M_{\rm Jup}$)} 
&\colhead{($M_{\rm Jup}$)} 
&\colhead{($M_{\rm Jup}$)} 
} 
\startdata \\ 
\multicolumn{7}{c}{Gas Detections} \\ 
\hline \\ 
540 & 1204 $\pm$ 85 & 276 $\pm$ 54 & $<78$ & 2.4 & 1.0 & 10.5 \\
1274 & 861 $\pm$ 88 & 326 $\pm$ 68 & $<48$ & 5.5 & 1.0 & 31.4 \\
1152 & 633 $\pm$ 82 & 314 $\pm$ 65 & $<60$ & 7.1 & 1.0 & 31.4 \\
1153 & 557 $\pm$ 57 & $<99.0$ & $<72$ & ... & ... & 1.0 \\
818 & 514 $\pm$ 58 & $<108.0$ & $<81$ & ... & ... & 1.0 \\
1075 & 165 $\pm$ 33 & $<93.0$ & $<66$ & ... & ... & 10.5 \\
\hline \\ 
\multicolumn{7}{c}{Gas Non-detections} \\ 
\hline \\ 
1036 & $<72.0$ & $<81.0$ & $<57$ & ... & ... & 3.1 \\
1050 & $<72.0$ & $<81.0$ & $<60$ & ... & ... & 3.1 \\
1154 & $<69.0$ & $<75.0$ & $<57$ & ... & ... & 3.1 \\
1155 & $<69.0$ & $<78.0$ & $<57$ & ... & ... & 3.1 \\
1156 & $<72.0$ & $<84.0$ & $<60$ & ... & ... & 3.1 \\
1182 & $<69.0$ & $<78.0$ & $<60$ & ... & ... & 3.1 \\
1193 & $<72.0$ & $<81.0$ & $<60$ & ... & ... & 3.1 \\
1230 & $<72.0$ & $<81.0$ & $<57$ & ... & ... & 3.1 \\
1248 & $<72.0$ & $<81.0$ & $<60$ & ... & ... & 3.1 \\
1260 & $<69.0$ & $<78.0$ & $<60$ & ... & ... & 3.1
\enddata 
\tablenotetext{}{(This table is available online in its entirety in machine-readable form.)} 
\end{deluxetable*} 
\capstartfalse

\subsection{CO Line Emission \label{sec-line}}

To search for objects exhibiting significant line emission, we first extract the $^{12}$CO spectrum for each source. When creating the spectrum, we use 1~km~s$^{-1}$ velocity sampling and measure fluxes in each channel using a circular aperture 0\farcs30 in radius and centered on the continuum emission (for detections) or the expected stellar position (for non-detections). We measure the image rms using a 4\arcsec--9\arcsec radius annulus centered on the fitted or expected source position. Candidate detections are identified as those with emission exceeding 3$\times$ the rms in multiple nearby channels within 0--25~km~s$^{-1}$ (LSRK frame), which covers the range of RVs found for $\sigma$ Orionis members \cite[e.g.,][]{2006MNRAS.371L...6J}. 

For each candidate detection, we create zero-moment maps by integrating across the velocity range where the emission exceeds the noise. The integrated flux ($F_{\rm 12CO}$) is then measured using circular aperture photometry, where the aperture radius for each source is determined by a curve-of-growth method in which successively larger apertures are applied until the flux stabilizes to within errors. Uncertainties ($E_{\rm 12CO}$) are estimated by taking the standard deviation of the fluxes measured within the same sized aperture placed randomly within the field of view but away from the source. We consider sources as detections when $F_{\rm 12CO}>4 \times E_{\rm 12CO}$. We adopt this high detection threshold because, as pointed out in \cite{2016ApJ...827..142B}, this procedure selects both the velocity range and aperture size that maximize the signal, thus can produce false detections at lower significance levels. We detect only six sources in $^{12}$CO using this procedure. For these sources, we also search for $^{13}$CO and C$^{18}$O emission using the same velocity range and aperture photometry method as for $^{12}$CO; we detect three of these sources in $^{13}$CO and none in C$^{18}$O.

For sources with no significant line emission found using the above procedure, we create zero-moment maps by integrating across the channels $\pm$1~km~s$^{-1}$ from their known RVs, when available in the literature \citep{2008MNRAS.385.2210M,2008A&A...488..167S}. For sources with unknown RVs, we integrate around the average value for $\sigma$~Orionis members with known RVs. \cite{2006MNRAS.371L...6J} showed that $\sigma$~Orionis members are divided into two kinematically distinct subgroups differentiated by their RVs (Group~1 and 2, by their convention). However, the region is dominated by Group~2 sources at $\delta < -02$:18:00, where all our gas non-detections with unknown RVs are located. Thus we adopt the average RV of Group~2 (13~km~s$^{-1}$ in the LSRK frame) when creating zero-moment maps for our gas non-detections with unknown RVs. We measure $^{12}$CO, $^{13}$CO, and C$^{18}$O integrated fluxes from these zero-moment maps using the aforementioned aperture photometry method, but with an aperture size fixed to the beam size. We found no additional detections, thus took upper limits as 3$\times$ the image rms.

\begin{figure}
\begin{centering}
\includegraphics[width=8.7cm]{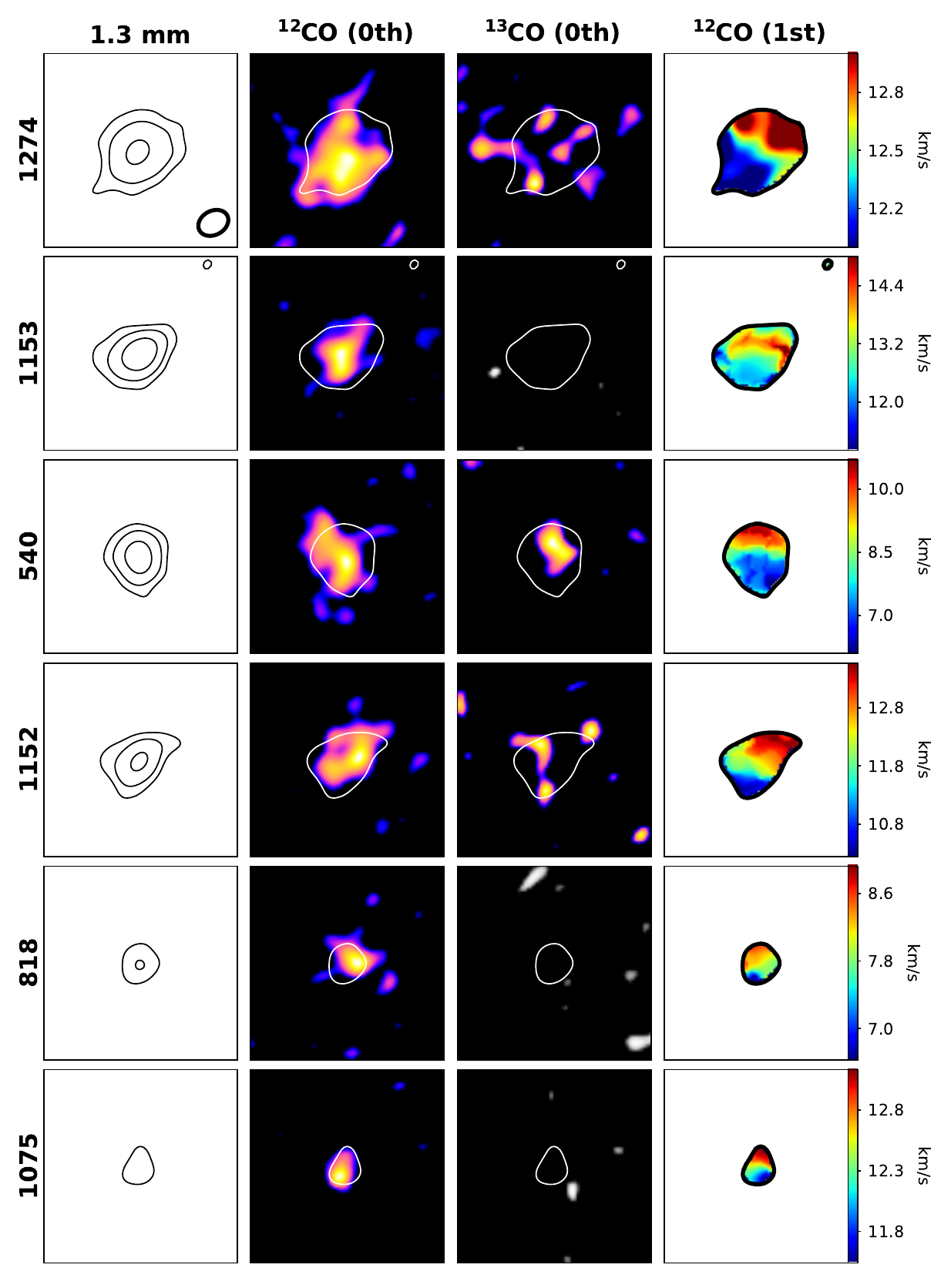}
\caption{\small The six sources in our sample detected in CO (Section~\ref{sec-line}). The first column shows the 1.33~mm continuum emission in 4$\sigma$, 10$\sigma$, and 25$\sigma$ contours. The second and third columns show the $^{12}$CO and $^{13}$CO zero-moment maps with 4$\sigma$ continuum contours. The last column shows the $^{12}$CO first-moment maps within 4$\sigma$ continuum contours. Images are 2\arcsec$\times$2\arcsec and the typical beam size is given in the first panel.}
\label{fig-gas}
\end{centering}
\end{figure}

Table~\ref{tab-gas} gives our integrated line fluxes or upper limits. Of the 92 targets, only six are detected in $^{12}$CO, three are detected in $^{13}$CO, and none are detected in C$^{18}$O with $>4\sigma$ significance. All sources detected in $^{12}$CO are detected in the continuum, and all sources detected in $^{13}$CO are detected in $^{12}$CO. The zero- and first-moment maps of the gas detections are shown in Figure~\ref{fig-gas}.

 \capstartfalse
\capstartfalse

     
\section{PROPERTIES OF $\sigma$ ORIONIS DISKS}
\label{sec-analysis}
 
\capstartfalse
\begin{figure*}
\begin{centering}
\includegraphics[width=18.2cm]{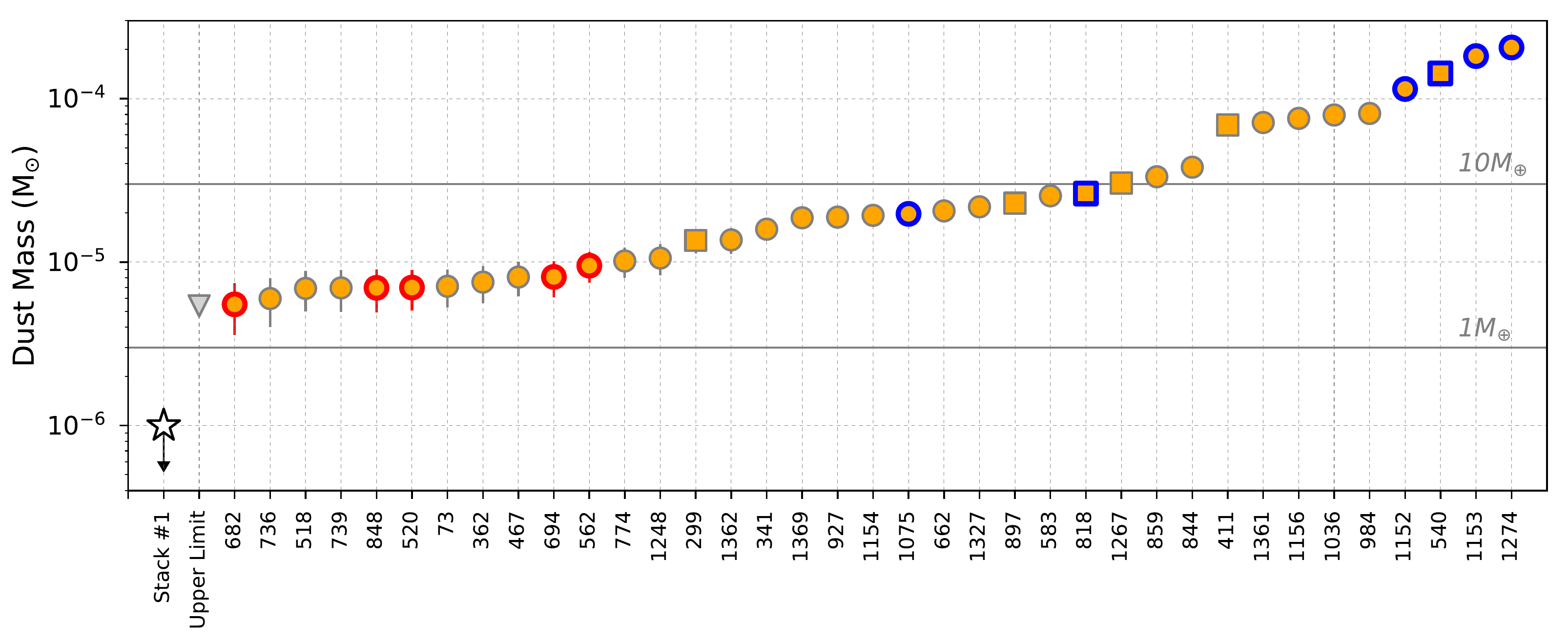}
\caption{\small Dust masses for the 37 continuum-detected sources in our $\sigma$~Orionis ALMA sample. Dust masses are from Table~\ref{tab-cont} and error bars include the 10\% absolute flux calibration uncertainty (Section~\ref{sec-observations}). The downward-facing triangle is the typical 3$\sigma$ upper limit for individual continuum non-detections, while the star shows the constraint on their average dust mass from the stacked continuum non-detections (``Stack \#1" in Section~\ref{sec-stacks}). Squares indicate TDs identified by their SEDs (Section~\ref{sec-TD}). Sources outlined in blue are also detected in $^{12}$CO (Section~\ref{sec-gas}) and sources outlined in red are $\lesssim0.5$~pc from the central OB system (Section~\ref{sec-photoevap}); sources closest to the central OB system have the lowest dust masses, while sources with the highest dust masses tend to also be detected in CO.}
\label{fig-dustmass}
\end{centering}
\end{figure*}
\capstartfalse

 \subsection{Dust Masses \label{sec-dust}}
 
Because dust emission at (sub-)mm wavelengths can be optically thin, the bulk dust mass of a disk ($M_{\rm dust}$) can be estimated from its (sub-)mm continuum emission at a given wavelength ($F_{\nu}$), as shown in \cite{1983QJRAS..24..267H}:

\begin{equation}
M_{\rm dust}=\frac{F_{\nu}d^{2}}{\kappa_{\nu}B_{\nu}(T_{\rm dust})}.
\label{eqn-mass}
\end{equation}

where $B_{\nu}(T_{\rm dust})$ is the Planck function for a characteristic dust temperature of $T_{\rm dust}=20$~K (the median for Taurus disks; \citealt{2005ApJ...631.1134A}). We take the dust grain opacity, $\kappa_{\nu}$, as 10 cm$^{2}$ g$^{-1}$ at 1000 GHz and use an opacity power-law index of $\beta=1$ \citep{1990AJ.....99..924B}. The source distance, $d$, is taken as 385~pc based on the updated parallax of the $\sigma$~Ori triple system \citep{2016AJ....152..213S}. Equation~\ref{eqn-mass} can therefore be approximated as $M_{\rm dust}\approx1.34\times10^{-5}~F_{1.33{\rm mm}}$, where $F_{1.33{\rm mm}}$ is in mJy and $M_{\rm dust}$ is in $M_{\odot}$. 

We use this simplified approach, which assumes a single grain opacity and an isothermal disk temperature, to ease comparisons with other disk surveys (Sections~\ref{sec-slopes} \& \ref{sec-dustdist}). Moreover, although \cite{2013ApJ...771..129A} derived a $T_{\rm dust}=25 {\rm K} \times (L_{\ast}/L_{\odot})^{0.25}$ relation from two-dimensional continuum radiative transfer models, we use an isothermal disk temperature as more detailed modeling of resolved disks suggests that $T_{\rm dust}$ is independent of stellar parameters. In particular, Tazzari et al. (2017, submitted) fit continuum observations of 36 resolved disks in Lupus directly in the uv-plane to two-layer disk models by solving energy balance equations at each disk radius, finding no dependence of $T_{\rm dust}$ as a function of stellar parameters.

Table~\ref{tab-cont} presents our $M_{\rm dust}$ estimates, derived using Equation~\ref{eqn-mass} with our $F_{\rm 1.33mm}$ measurements (Section~\ref{sec-cont}). Figure~\ref{fig-dustmass} shows the continuum-detected disks in order of increasing $M_{\rm dust}$ as well as the typical 3$\sigma$ upper limit of $\sim$2.0~$M_{\oplus}$ ($F_{\rm 1.33mm}\sim0.45$~mJy). Only 4 disks have $M_{\rm dust}>30~M_{\oplus}$ ($F_{\rm 1.33mm}>6.7$~mJy), thus nearly all protoplanetary disks in $\sigma$~Orionis have dust masses well below the minimum mass of solids needed to form the planets in our Solar System \citep{1977Ap&SS..51..153W}. Moreover, only 11 disks have $M_{\rm dust}>10~M_{\oplus}$ ($F_{\rm 1.33mm}>2.2$~mJy), thus by $\sim$3--5~Myr most protoplanetary disks appear to lack sufficient dust to form giant planet cores. Note that significant amounts of solids may still exist in objects larger than a few cm in size, which do not produce detectable (sub-)mm emission.

 \subsection{Gas Masses \label{sec-gas}}

CO line emission can be used to roughly estimate bulk gas masses ($M_{\rm gas}$) independently from the dust, assuming simple CO chemistry and adopting an ISM-like CO/H$_2$ abundance. \cite{2014ApJ...788...59W} (hereafter WB14) used parametrized gas disk models to demonstrate that the majority of CO may exist in the warm molecular layer, where it is sufficiently warm to survive freeze-out onto the disk midplane as well as adequately shielded from UV radiation to avoid photodissociation (except for particularly cold or low-mass disks). The CO isotopologue lines, specifically $^{13}$CO and C$^{18}$O, are especially useful for constraining $M_{\rm gas}$ as their moderate-to-low optical depths mean that they trace the bulk gas content rather than the temperature profile of the disk. \cite{2016ApJ...828...46A} and \cite{2016arXiv161201538M} have used these CO isotopologues lines to estimate $M_{\rm gas}$ for protoplanetary disks in the Lupus clouds that have been surveyed by ALMA. 

\begin{figure}
\begin{centering}
\includegraphics[width=8.5cm]{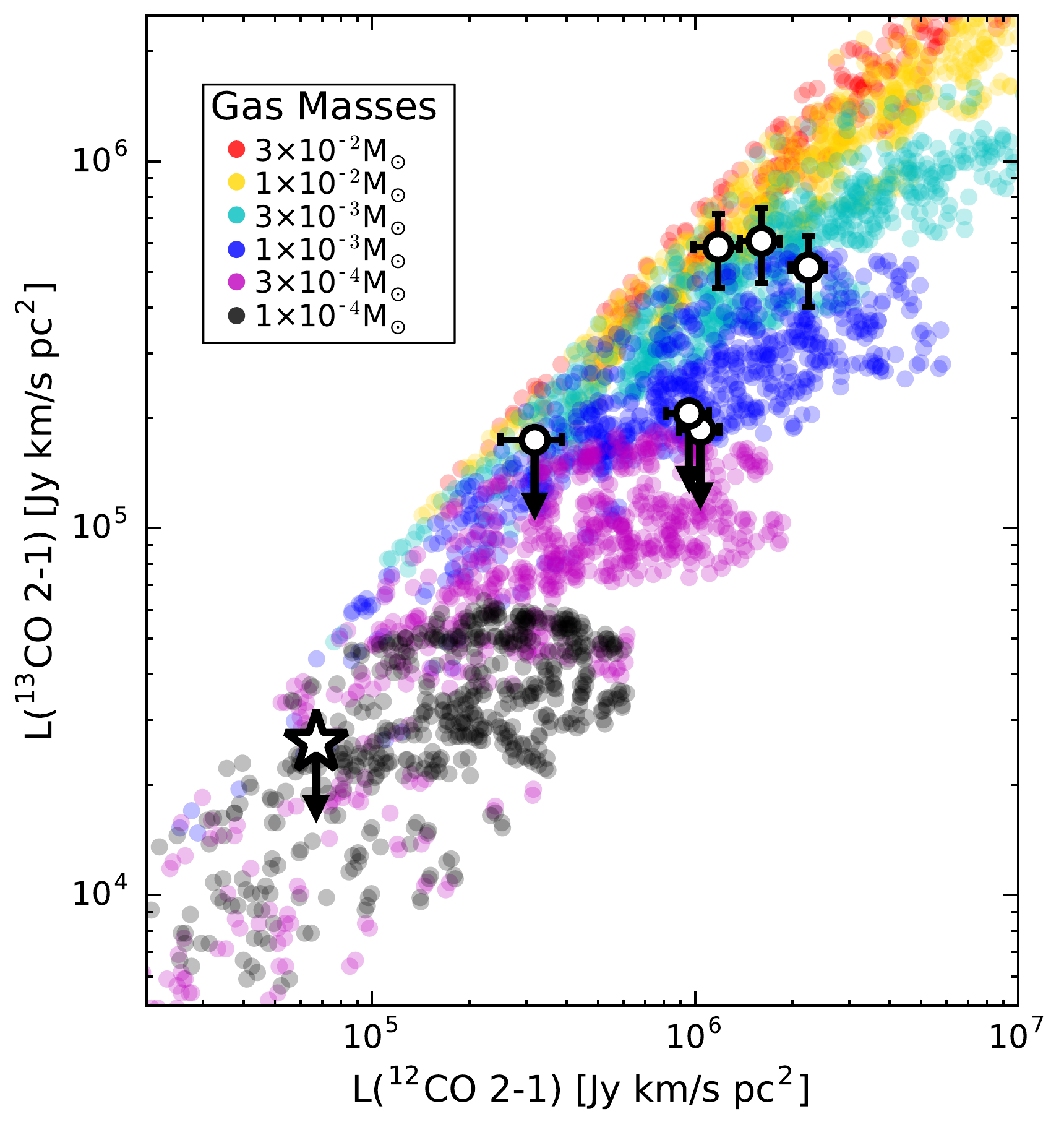}
\caption{\small $^{12}$CO and $^{13}$CO $J=1$--2 line luminosities for determining gas masses (Section~\ref{sec-gas}). Colored points show the WB14 model grid color-coded by gas mass. The three disks with both lines detected are plotted as white circles, and the three disks with only $^{12}$CO detections are plotted as white circles with arrows indicating 3$\sigma$ upper limits on $^{13}$CO. Error bars account for the statistical errors given in Table~\ref{tab-gas} as well as the 10\% absolute flux calibration error. The star shows the location of ``Stack \#2" (Section~\ref{sec-stacks}) where the error bars are smaller than the symbol.}
\label{fig-gasmass}
\end{centering}
\end{figure}

Unfortunately, because we find no C$^{18}$O detections in $\sigma$~Orionis, we cannot use the same combination of CO isotopologue lines to estimate $M_{\rm gas}$ in this region. However, we can still place rough constraints on $M_{\rm gas}$ by comparing our measured $^{12}$CO and $^{13}$CO line luminosities or upper limits to the WB14 model grid. The uncertainties on $M_{\rm gas}$ are larger for this line combination because $^{12}$CO is optically thick and therefore more sensitive to other disk parameters, such as the temperature profile. Nevertheless, $M_{\rm gas}$ can still be estimated using this method because the combination of integrated line fluxes still primarily depends on bulk gas mass rather than these other disk parameters (see the parameter exploration described in WB14 as well as the separation of gas masses in Figure~\ref{fig-gasmass}). 

Figure~\ref{fig-gasmass} compares the WB14 model grid to our measured $^{12}$CO and $^{13}$CO line luminosities for sources detected in at least one of these lines. Table~\ref{tab-gas} provides our $M_{\rm gas}$ constraints derived from the WB14 model grid. To estimate $M_{\rm gas}$ for the three sources detected in both $^{12}$CO and $^{13}$CO (540, 1274, 1152), we calculate the mean (in log space) of the WB14 model grid points consistent with our measured line luminosities and their associated errors; these values span 2--7~$M_{\rm Jup}$. We also set upper ($M_{\rm gas,max}$) and lower ($M_{\rm gas,min}$) limits based on the maximum and minimum WB14 model grid points consistent with the data, respectively. For the three sources with $^{12}$CO detections and $^{13}$CO upper limits (1153, 818, 1075) we provide only $M_{\rm gas,max}$, since the lower bound is constrained by the limits of the WB14 model grid, and at these very low masses photodissociation becomes important and self-shielding needs to be taken into account. For the 86 sources undetected in both lines, we give only $M_{\rm gas,max}$, which is set by the maximum WB14 model grid point consistent with the upper limits on both lines.

\capstartfalse 
\begin{deluxetable}{lccc} 
\tabletypesize{\footnotesize} 
\centering 
\tablewidth{240pt} 
\tablecaption{Gas Properties derived from \cite{2016A&A...594A..85M} models \label{tab-M16}} 
\tablecolumns{4}  
\tablehead{ 
 \colhead{Source} 
&\colhead{$M_{\rm gas}$ ($M_{\rm Jup}$)} 
&\colhead{Gas-to-dust ratio} 
&\colhead{Line} 
} 
\startdata 
540        &  $<5.8$ & $<41$ & C$^{18}$O \\
1274      &  $<3.1$ & $<15$ &  C$^{18}$O \\
1152      &  $<4.2$ & $<37$ &  C$^{18}$O \\
1153      &  $<0.3$ & $<2$ &  $^{13}$CO \\
818        &  $<0.4$ & $<15$ &  $^{13}$CO \\
1075      &  $<0.3$ & $<15$ &  $^{13}$CO
\enddata 
\end{deluxetable} 
\capstartfalse 

\capstartfalse
\capstartfalse

Alternatively, one can estimate $M_{\rm gas}$ by employing the grid of physical-chemical models presented in \cite{2016A&A...594A..85M} (hereafter M16). Using the code DALI \citep{2012A&A...541A..91B}, M16 investigated a range of realistic disk and stellar parameters, calculating self-consistently the thermal and chemical structure of the disk up to an age of 1~Myr. As shown by \cite{2016arXiv161201538M} for Lupus disks, the simulated line fluxes can be used to derive $M_{\rm gas}$ from individual $^{13}$CO or C$^{18}$O isotopologue line observations, assuming an abundance of volatile carbon. This is because the medians of the simulated line luminosities can be fit by simple functions of $M_{\rm gas}$ (see Equation~2 in M16). Namely, for low-mass disks the dependence of the line luminosities on $M_{\rm gas}$ is linear, while the trend becomes logarithmic for more massive disks due to the line emission becoming optically thick. Thus, the observed $^{13}$CO or C$^{18}$O line luminosities directly trace $M_{\rm gas}$ only if they fall in the linear dependence regime found in M16.  

Applying the methodology of M16 to our $\sigma$~Orionis sample, we find that the $^{13}$CO upper limits for the three sources detected only in $^{12}$CO (1153, 818, 1075) are in the linear dependence regime and thus provide upper limits on $M_{\rm gas}$. Although the three brightest disks (540, 1274, 1152) have $^{13}$CO line luminosities that fall in the logarithmic dependence regime, their C$^{18}$O upper limits can still provide constraints on $M_{\rm gas}$. The derived $M_{\rm gas}$ upper limits and the CO isotopologue lines used for the calculations are reported in Table~\ref{tab-M16}. These results place stronger constraints on $M_{\rm gas}$ than the WB14 model grid for all six sources detected in CO.

There are several caveats to our derived gas masses. Namely, these gas masses depend inversely on the assumed molecular CO abundance of ${\rm [CO]/[H_{2}]}=10^{-4}$ in the case of WB14, and the assumed volatile carbon abundance of ${\rm [C]/[H]}=1.35\times10^{-4}$ and chemical age of 1~Myr in the case of M16. Additionally, both WB14 and M16 assume an isotopologue ratio of ${\rm [CO]/[^{13}CO]}=70$.  The WB14 ${\rm [CO]/[H_{2}]}$ abundance is consistent with those measured in molecular clouds \citep{1982ApJ...262..590F,1994ApJ...428L..69L,2013MNRAS.431.1296R,2014A&A...564A..68S} as well as with a direct measurement in a disk \citep{2014ApJ...794..160F}. However, the strong HD \citep{2013Natur.493..644B} but weak C$^{18}$O emission toward the TW Hydra disk has been interpreted as resulting from significant carbon depletion of up to two orders of magnitude in this system \citep{2013ApJ...776L..38F,2016A&A...588A.108K,2016ApJ...823...91S}. 

If carbon depletion (rather than gas depletion) is the true cause of weak CO emission, the responsible physical mechanisms are not yet established. One hypothesis is that gas-phase reactions initiated by X-ray and cosmic ray ionization of He produce He$^+$ atoms that react with gaseous CO to gradually extract carbon, which is then processed into more complex molecules that freeze onto cold dust grains at higher temperatures than CO \citep{1997ApJ...486L..51A,2012A&A...541A..91B, 2013ApJ...776L..38F, 2014FaDi..168...61B, 2016A&A...588A.108K, 2016ApJ...822...53Y}. Alternatively, CO can be turned into more complex organics such as CH$_3$OH, or into CO$_2$ and CH$_4$, via ice chemistry reactions \cite[e.g., see Figure 3c in][]{2016A&A...595A..83E}; these reactions have typical timescales of a few Myr (depending on the ionization rate) and thus could be more significant in older systems. Finally, volatile carbon may be locked up in large icy bodies in the disk midplane \citep{2010A&A...521L..33B,2013A&A...552A.137R,2016A&A...588A.112G}. These large pebbles cannot diffuse upward and thus would no longer participate in gas-phase chemistry \citep{2015ApJ...807L..32D,2016A&A...588A.108K}. Such a process would ``dry out" the CO from the warm molecular layer, analogous to what is proposed to explain the under-abundance of gas-phase water in disk atmospheres \citep{2010A&A...521L..33B,2011Sci...334..338H}. If any of these mechanisms significantly depletes carbon in the disk, our derived gas masses would be underestimated.

\subsection{Stacking Analysis \label{sec-stacks}}

We perform a stacking analysis to constrain the average dust and gas masses of the individually undetected sources in our sample. To stack the images, we average them in the image plane after centering them on their expected source locations. We then search for emission using the aperture photometry method described in Section~\ref{sec-line}.

We first stack the 55 sources undetected in the continuum (``Stack \#1"), but do not find a significant mean signal in the continuum or any of the CO lines. The measured continuum mean signal is 0.05$\pm$0.03~mJy (1.7$\sigma$). We confirm this non-detection by calculating the mean and standard error on the mean from the continuum fluxes reported in Table~\ref{tab-cont}, which similarly gives 0.03$\pm$0.02~mJy (1.5$\sigma$). This provides a 3$\sigma$ upper limit on the average dust mass of individually undetected continuum sources of 0.4~$M_{\oplus}$, which is 5$\times$ lower than the smallest dust mass among the continuum detected sources in $\sigma$~Orionis (see Figure~\ref{fig-dustmass}). This striking difference in the dust masses of detected and undetected continuum sources was also seen in an ALMA survey of Lupus disks \cite[see Figure~3 in][]{2016ApJ...828...46A} and further supports theoretical models that predict protoplanetary disks dispersing rapidly once disk clearing begins \cite[e.g., see review in][]{2014prpl.conf..475A}.

We also stack the 31 sources that are detected in the continuum but undetected in $^{12}$CO (``Stack \#2"), finding a mean continuum signal of $2.29\pm0.09$~mJy as well as a significant mean $^{12}$CO signal of $36\pm8$~mJy~km~s$^{-1}$ (4.5$\sigma$). No emission is detected in the $^{13}$CO or C$^{18}$O lines with 3$\sigma$ upper limits of 14 and 11~mJy~km~s$^{-1}$, respectively. The continuum flux corresponds to a dust mass of $\sim$10~$M_{\oplus}$, while the $^{12}$CO detection and $^{13}$CO upper limit correspond to a gas mass $<1.0~M_{\rm Jup}$ using the WB14 model grid (the $^{13}$CO and C$^{18}$O upper limits also correspond to a gas mass $<1.0~M_{\rm Jup}$). This gives an average gas-to-dust ratio of $<30$ for sources detected in the continuum but not in CO, assuming standard CO abundances.

\subsection{Transition Disks \label{sec-TD}}

Transition disks (TDs) are protoplanetary disks with large inner cavities in their dust distributions \cite[see review in][]{2014prpl.conf..497E}. TDs can be identified by resolved (sub-)mm images, or by the mid-IR deficits in their SEDs, which indicate a lack of warm micron-sized dust grains close to the central star. Eight $\sigma$~Orionis members (1268, 1267, 908, 897, 818, 540, 411, 299) have been identified as TDs based on their SEDs \citep{2007ApJ...662.1067H,2016ApJ...829...38M} and all of these sources are included in our ALMA sample.

Six of these TDs (1267, 897, 818, 540, 411, 299) were detected by our ALMA continuum observations, and two of the continuum-detected sources (818, 540) were also detected in CO. Our ALMA observations did not resolve any dust cavities (see Figure~\ref{fig-cont}), although this is not particularly surprising given the large beam size of our observations ($\sim$120$\times$95~AU at 385~pc; Section~\ref{sec-observations}). 

Thus 16\% (6/37) of our ALMA continuum detections are TDs (see Figure~\ref{fig-dustmass}), which is interestingly similar to the 19\% fraction found in Lupus \citep{2016ApJ...828...46A} and consistent with previous findings that TDs tend to be among the brightest disks in a given star-forming region \citep{2011ApJ...732...42A,2016ApJ...828...46A}. Moreover, the TD fraction across the entire protoplanetary disk population in $\sigma$~Orionis is 8\% (8/92), which is consistent with TD fractions found in other star-forming regions \cite[see Figure 11 in ][]{2014prpl.conf..497E}.


\section{DISCUSSION}
\label{sec-discussion}

\subsection{External UV Photoevaporation from $\sigma$ Ori  \label{sec-photoevap}}

\capstartfalse
\begin{figure*}
\begin{centering}
\includegraphics[width=18.cm]{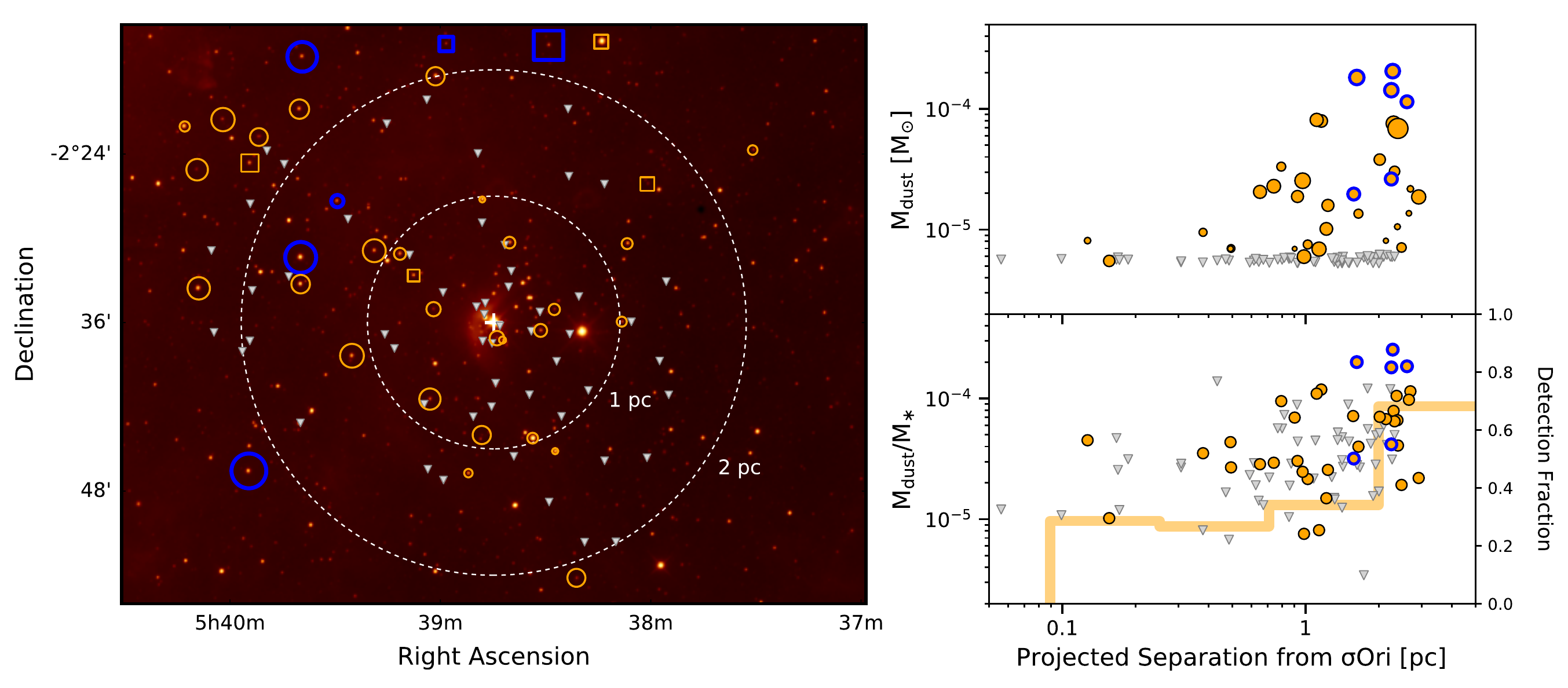}
\caption{\small {\it Left}: map of $\sigma$ Orionis with our ALMA continuum detections circled in orange and gas detections circled in blue; the sizes of the circles scale with the ratio of disk dust mass to stellar mass ($M_{\rm dust}/M_{\ast}$) and squares indicate TDs. Non-detections are shown by gray triangles. The central OB system, $\sigma$ Ori, is marked by the white cross and the dashed white circles show radial distances of 1~pc and 2~pc. Notably, the sources with detectable gas emission are among the furthest from $\sigma$~Ori. {\it Top right}: disk dust mass ($M_{\rm dust}$) as a function of projected separation from $\sigma$~Ori, where orange points are continuum detections and gray triangles are 3$\sigma$ upper limits. There is a clear decline in $M_{\rm dust}$ at smaller separations from the central OB system, and massive disks ($M_{\rm dust}\gtrsim3~M_{\odot}$) are missing within $\sim$0.5~pc of $\sigma$~Ori. {\it Bottom right}: $M_{\rm dust}/M_{\ast}$ as a function of projected separation from $\sigma$~Ori, illustrating that the declining trend still holds even after correcting for the $M_{\rm dust}$--$M_{\ast}$ relation (Section~\ref{sec-slopes}). Our ALMA continuum detection fraction, shown by the thick orange line, also stays relatively constant until $\sim$2~pc, after which it doubles.}
\label{fig-ob}
\end{centering}
\end{figure*}
\capstartfalse

OB associations host at their centers very massive stars with O and B type spectral classifications, surrounded by populations of several hundred to thousands of lower-mass stars. The massive OB stars emit large numbers of extreme-ultraviolet (EUV; $h\nu>13.6$~eV) and far-ultraviolet (FUV; 6~eV$~<h\nu<~$13.6~eV) photons, which can heat and photoevaporate circumstellar disks around the nearby low-mass stars. The resulting mass-loss rate for a given disk depends on the relative strengths of the incident FUV and EUV fluxes, and thus should depend on the distance of the source from the central OB stars \cite[e.g.,][]{1998ApJ...499..758J,1999ApJ...515..669S}. 

\cite{2014ApJ...784...82M} used ALMA to search for observational evidence of these external photoevaporation effects on proplyds in the young ($\sim$1--2~Myr) Orion Nebula Cluster (ONC). They found a clear drop in disk mass at small projected separations from the central O6 star, $\theta^{1}$~Ori~C. Namely, there was a lack of massive ($M_{\rm dust}\gtrsim9~M_\oplus$) disks within $\sim$0.03 pc of $\theta^{1}$~Ori~C, where models predict that EUV emission dominates the radiation field \citep{1998ApJ...499..758J}. At larger separations of $\sim$0.03--0.3~pc, where less energetic FUV emission is expected to dominate \citep{2004ApJ...611..360A}, they found a range of disk masses representative of typical low-mass star-forming regions, indicating that the lower mass-loss rates in FUV-dominated regions can preserve disk masses for up to a couple Myr. \cite{2014ApJ...784...82M} concluded that planet formation is likely inhibited for disks in the inner-most EUV-dominated regions of OB associations due to higher mass-loss rates, while disks in the FUV-dominated regions and beyond are relatively unaffected with planet formation proceeding as in isolated disks.

\cite{2015ApJ...802...77M} looked for similar effects in the very young NGC~2024 region, which at $\sim$0.5~Myr old \cite[e.g.,][]{2006ApJ...646.1215L} hosts the massive star IRS~2b of O8--B2 spectral type \citep{2003A&A...404..249B} as well as several hundred YSOs still heavily embedded in molecular cloud material. \cite{2015ApJ...802...77M} could not identify a distance-dependent disk mass distribution in NGC~2024, and instead found several massive ($M_{\rm dust}\gtrsim17~M_\oplus$) disks located $<0.01$~pc from IRS~2b. They argued that this could be an evolutionary effect: the extremely young age of NGC~2024 simply means that processes like external photoevaporation have not yet had time to significantly reduce disk masses. Alternatively, they suggested this could be an environmental outcome: the significant cloud material in NGC~2024 may efficiently absorb the high-energy photons from IRS~2b, or the later spectral type of the star (compared to $\theta^{1}$~Ori~C in the ONC) means that it does not produce sufficient photoionizing radiation.

Here we search for evidence of external photoevaporation in $\sigma$~Orionis, an OB association whose central trapezium system, $\sigma$~Ori, contains a massive O9 star. $\sigma$~Orionis is an interesting target for studying external photoevaporation, as its lack of cloud material and older age may both enhance the observable effects of external photoevaporation. Figure~\ref{fig-ob} (upper right panel) plots disk dust mass as a function of projected separation from $\sigma$~Ori ($\alpha$ = 05:38:44.779, $\delta=-$02:36:00.11). Similar to the ONC, we find a lack of massive ($M_{\rm dust}\gtrsim3~M_\oplus$) disks close to the central OB system; however, the drop in occurrence is seen at $\sim$0.5~pc---a much larger projected distance compared to the $\sim$0.03~pc limit found for the ONC \citep{2014ApJ...784...82M}. Moreover, beyond $\sim$0.5~pc, we see a smooth distance-dependent dust mass distribution that extends out to several parsecs. We note that the smaller dust masses and larger projected distances found in $\sigma$~Orionis (compared to those found in the ONC) are both influenced by the older age of the region. Namely, dust mass distributions are known to decline with cluster age (Section~\ref{sec-dustdist}) and typical intra-cluster velocity dispersions of several km~s$^{-1}$ can result in cluster expansions of several parsecs by the age of $\sigma$~Orionis. 

One concern is that, due to the $M_{\rm dust}$--$M_{\ast}$ relation (Section~\ref{sec-slopes}), mass segregation in clusters could produce these observed trends, if the least massive stars are preferentially located closer to the cluster centers. \cite{2014ApJ...784...82M} could not test this in the ONC, as the nature of proplyds complicates any estimates of stellar mass. Because we can estimate stellar masses in $\sigma$~Orionis (Section~\ref{sec-sample}), we also show in Figure~\ref{fig-ob} (lower right panel) the ratio of disk dust mass to stellar mass ($M_{\rm dust}/M_{\ast}$) as a function of projected separation, confirming that the distance-dependent trend still holds even when accounting for stellar mass differences. Moreover, Figure~\ref{fig-ob} shows our ALMA continuum detection fraction, illustrating a relatively constant detection rate of $\sim$30\% out to $\sim$2~pc, after which the detection fraction more than doubles to $\sim$70\%. \cite{2007ApJ...662.1067H} did not find a similar change in detection fraction with their {\it Spitzer} survey of $\sigma$~Orionis disks (see their Figure 16), however this may be because external photoevaporation does not remove the inner (i.e., more gravitationally bound) disk regions probed by {\it Spitzer}.

Interestingly, we also find that the CO detections in our sample (blue circles in Figure~\ref{fig-ob}) only exist in the outer regions of the cluster. This is qualitatively consistent with the picture of external photoevaporation: for typical disks, the gas is generally more extended than the dust, and therefore less tightly bound to the star, making the gas more susceptible to external photoevaporation. However, our gas sample is small and \cite{2014ApJ...784...82M} were unable to reliably detect gas in ONC disks due to cloud confusion, making it important to confirm our finding with surveys of other OB associations. If external photoevaporation does have a more significant effect on gas relative to dust, this would impact the types of planets that can form in OB associations.

Some evidence for external photoevaporation has been previously found for $\sigma$ Orionis disks. \cite{2009A&A...495L..13R} detected strong optical forbidden emission lines from SO~587, which they interpreted as an externally driven photoevaporative flow due to the very low stellar mass accretion rate for this source, the profile shapes and luminosities of the forbidden emission lines, and the small projected separation ($\sim$0.3~pc) of the disk from $\sigma$~Ori (we did not detect this source with our ALMA observations). Additionally, \cite{2016ApJ...829...38M} fit irradiated accretion disk models to the SEDs of 18 sources in $\sigma$~Orionis to show decreased disk masses and sizes when compared to those in the younger ONC. They interpreted this as evidence for external photoevaporation, however their results were uncertain due to various model assumptions (e.g., constant $\alpha=0.01$) as well as the comparison of disk properties derived from disparate methods (e.g., they compared $\sigma$~Orionis disk radii derived from SED modeling, which probes the dust disk, to ONC disk radii derived from {\it Hubble} imagery, which probes the gas disk). Our ALMA observations therefore provide the clearest evidence to date that external photoevaporation is affecting disk masses throughout the $\sigma$~Orionis region. 

Our findings also indicate that FUV (not just EUV) emission from OB stars is an important driver of external photoevaporation. Assuming a typical O9V FUV luminosity of $\log{(L_{\rm FUV}/L_\odot)}=4.5$ for $\sigma$~Ori, the geometrically diluted FUV flux within the region can be expressed as $\sim 8000 (d/{\rm pc})^{-2}\,G_0$, where $d$ represents the distance from the photoevaporative source in parsecs and $G_0 = 1.6\times10^{-3}$\,erg\,cm$^{-2}$\,s$^{-1}$ \citep{1968BAN....19..421H}. We note that although $\sigma$~Ori is a triple system, the FUV flux is usually dominated by the most massive star in the cluster \citep{2008ApJ...675.1361F,2011PASP..123...14H}. In this simple calculation, we also do not consider any extinction due to intra-cluster dust, which is observed to be at low densities in $\sigma$~Orionis \citep{2008hsf1.book..732W}, unlike in the ONC and NGC~2024. Figure~\ref{fig-ob} shows that external photoevaporation is affecting disk masses out to at least $\sim$2~pc, which when combined with the above equation corresponds to FUV fluxes $\gtrsim2000\,G_0$.

This supports recent observations that suggest even moderate FUV fluxes can drive significant mass loss. \cite{2016ApJ...826L..15K} observed 7 proplyds near a B star in NGC 1997, finding high mass-loss rates for an FUV flux of only $\sim$$3000\,G_0$. \cite{haworth2017} also showed that the outer disk of IM~Lup may be undergoing photoevaporation from an FUV flux of just $\sim$$4\,G_0$, where the high mass-loss rate can be explained by the large size of the disk \citep{2016ApJ...832..110C}, which causes gas in the outer regions to be only weakly gravitationally bound to the central star. Futhermore, \cite{2016arXiv160501773G} found that disk frequency (as probed by near-IR excess) declines with smaller projected separation from the OB stars in Cygnus OB2 for FUV fluxes $\gtrsim1000\,G_0$, similar to what we calculated for $\sigma$~Orionis. 

Together, these observations support recent theoretical findings by \cite{2016MNRAS.457.3593F} and \cite{2016MNRAS.463.3616H}, who predicted high mass-loss rates due to external photoevaporation for FUV fluxes $<3000\,G_0$. Moreover, these slow photoevaporative winds should be much more effective at removing gas and small ($\lesssim1~\mu$m) dust particles compared to larger ($\gtrsim1$~mm) solids \citep{2016MNRAS.457.3593F}, which may help to explain our lack of gas detections at projected distances $\lesssim1.5$~pc from $\sigma$~Ori.

\capstartfalse
\begin{figure*}
\begin{centering}
\includegraphics[width=18.2cm]{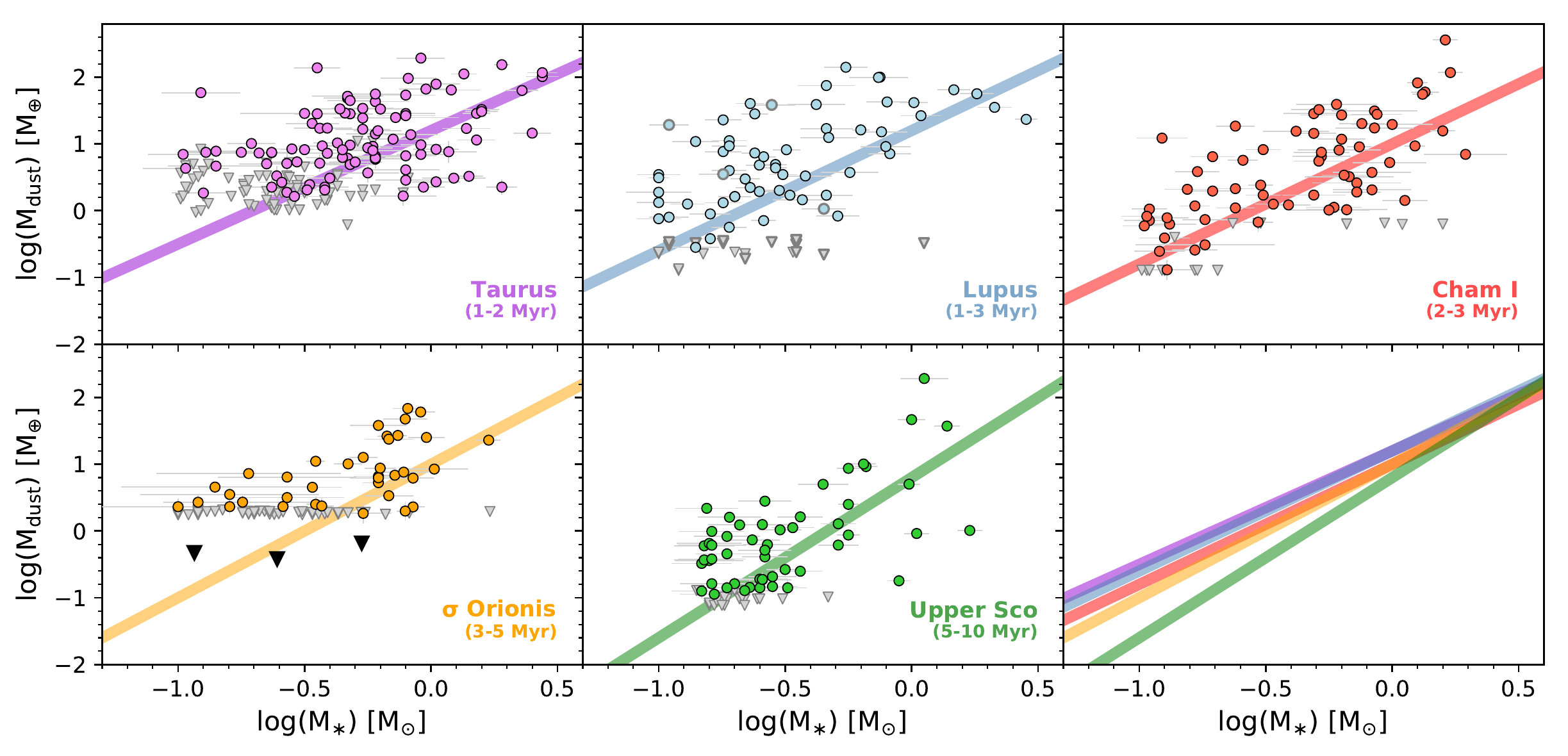}
\caption{\small Disk dust mass ($M_{\rm dust}$) as a function of stellar mass ($M_{\ast}$) for disk populations in five star-forming regions with ages spanning the disk dispersal timescale ($\sim$1--10~Myr). Colored circles are (sub-)mm continuum detections and gray triangles are 3$\sigma$ upper limits. For $\sigma$~Orionis, the black triangles indicate 3$\sigma$ upper limits from stacks of the non-detections in three stellar mass bins. For Lupus, the 20 sources with unknown stellar masses that were included in the analysis via an MC method \cite[see][]{2016ApJ...828...46A} are given representative values and identified by thick gray outlines. For each region, the solid lines show the Bayesian linear regression fits to the data, which take into account upper limits, intrinsic scatter, and measurement errors on both axes \citep{2007ApJ...665.1489K}. The lower right panel compares the fits in all five regions, illustrating the $\sim$1~dex difference in $M_{\rm dust}$ between the youngest and oldest regions at low stellar masses, and the convergence in $M_{\rm dust}$ at high stellar masses.}
\label{fig-mdust}
\end{centering}
\end{figure*}
\capstartfalse

Finally, we showed in Section~\ref{sec-stacks} that the average dust mass of the undetected sources in $\sigma$~Orionis is at least $\sim$5$\times$ lower than the smallest dust mass among the continuum detections, implying that disk dispersal occurs on short timescales. The rapid dispersal of disks impinged by intermediate external FUV fluxes has been predicted by \citet{2007MNRAS.376.1350C} and later by \citet{2013ApJ...774....9A}. Their models combined estimated mass-loss rates from external FUV photoevaporation with viscous disk evolution to show that disks should be dispersed from the outside in on timescales much shorter than the expected disk lifetime. The typical lifetime of a viscous disk impinged with a $\sim$3000$\,G_0$ FUV flux was predicted to be roughly a few Myr, in agreement with our observations.

In summary, our observations indicate that external photoevaporation due to FUV emission from OB stars is significantly affecting disk evolution throughout the $\sigma$~Orionis cluster, although other disk evolution mechanisms are also clearly at play (Section~\ref{sec-slopes} \& \ref{sec-dustdist}). This additional depletion of dust and gas for disks in OB clusters should have implications for planet formation, and detailed theoretical studies will be needed to quantify the impacts for different planet types and to identify any relations to trends seen in the exoplanet population.

\subsection{$M_{\rm dust}$--$M_{\ast}$ Relation \label{sec-slopes}}

A clear correlation seen across protoplanetary disk populations is the positive relation between $M_{\rm dust}$ and $M_{\ast}$ (e.g., Figure~\ref{fig-mdust}). Evidence for the $M_{\rm dust}$--$M_{\ast}$ relation was first identified in pre-ALMA surveys of the Taurus star-forming region \citep{2000prpl.conf..559N,2013ApJ...771..129A} then later recovered with ALMA for other star-forming regions including the similarly aged Lupus clouds \citep{2016ApJ...828...46A} and Chamaeleon~I region \citep{2016ApJ...831..125P} as well as the more evolved Upper Sco association \citep{2016ApJ...827..142B}. 

The $M_{\rm dust}$--$M_{\ast}$ relation is pertinent to understanding planet formation because it tells us how disks are related to the properties of their host stars, which can then be tied to similar trends seen in the exoplanet population. For example, as previously noted by \cite{2013ApJ...771..129A}, the $M_{\rm dust}$--$M_{\ast}$ relation may fundamentally explain the correlation between giant planet frequency and host star mass \citep{2006ApJ...649..436E,2007ApJ...670..833J,2010ApJ...709..396B,2013A&A...549A.109B}, as the cores of giant planets theoretically form more efficiently both in higher-mass disks \cite[e.g.,][]{2008Sci...321..814T,2012A&A...541A..97M} and around higher-mass stars \cite[e.g.,][]{2008ApJ...673..502K}. Moreover, tracking the evolution of the $M_{\rm dust}$--$M_{\ast}$ relation with age can tell us whether disk evolution proceeds differently around low-mass stars compared to high-mass stars, which in turn can help to constrain the relative importance of different disk processes during planet formation \cite[e.g.,][]{2016ApJ...831..125P}.

However, parameterizing the $M_{\rm dust}$--$M_{\ast}$ relation is complicated by three main factors: measurement uncertainties on both variables, intrinsic scatter in the data, and upper limits. The procedure most often utilized in the disk survey literature is the Bayesian linear regression method of \cite{2007ApJ...665.1489K}, as it is capable of accounting for these three key factors simultaneously, unlike other linear regression methods (see \citealt{2016ApJ...831..125P} for a detailed discussion). For a given dataset, the \cite{2007ApJ...665.1489K} procedure fits a slope ($\beta$), intercept ($\alpha$), and intrinsic dispersion ($\delta$) with associated uncertainties on each parameter. 

Using this method, \cite{2016ApJ...828...46A} showed that the fitted slopes to the protoplanetary disk populations in the young ($\sim$1--3~Myr) Taurus and Lupus regions were consistent with each other, but both shallower than that of the older ($\sim$5--10~Myr) Upper Sco association. \cite{2016ApJ...831..125P} then showed that Chamaeleon~I had a slope consistent with the similarly aged Taurus and Lupus regions, further supporting a steepening of the $M_{\rm dust}$--$M_{\ast}$ relation with age. They also compared their results to theoretical models of grain growth, drift, and fragmentation to show that a steepening of the $M_{\rm dust}$--$M_{\ast}$ relation with age is consistent with the outer disk being in the fragmentation-limited regime. In this regime, grain sizes in the outer disk are limited by fragmenting collisions. When fragmentation sets the largest grain size, inward radial drift of dust occurs more rapidly around lower-mass stars, making their (sub-)mm continuum emission weaker and more compact with age compared to higher-mass stars.

Here we derive the $M_{\rm dust}$--$M_{\ast}$ relation for $\sigma$ Orionis disks, again using the Bayesian linear regression method of \cite{2007ApJ...665.1489K}. We only consider sources in our ALMA sample with $M_{\ast}\ge0.1~M_{\odot}$, so that we can compare our results to the relations derived for other star-forming regions (see below; this only removes 5 sources from our sample and does not affect the fit results). Using the $M_{\ast}$ and $M_{\rm dust}$ values in Table~\ref{tab-cont}, we derive a linear fit with $\alpha=1.0\pm0.2$, $\beta=2.0\pm0.4$, and $\delta=0.6\pm0.1$~dex, as shown in Figure~\ref{fig-mdust}. To help illustrate that the fit is properly accounting for the numerous non-detections, we also show 3$\sigma$ upper limits from stacks of the non-detections in several stellar mass bins. 

\capstartfalse 
\begin{deluxetable}{llccc} 
\tabletypesize{\footnotesize} 
\centering 
\tablewidth{240pt} 
\tablecaption{$M_{\rm dust}$--$M_{\ast}$ Bayesian Fit Parameters \label{tab-linmix}} 
\tablecolumns{5}  
\tablehead{ 
 \colhead{Region} 
&\colhead{Age (Myr)} 
&\colhead{$\alpha^{\ddagger}$} 
&\colhead{$\beta^{\ddagger}$} 
&\colhead{$\delta$} 
} 
\startdata 
Taurus                         &  1--2   & 1.2$\pm$0.1 & 1.7$\pm$0.2 & 0.7$\pm$0.1 \\
Lupus$^{\dagger}$      &  1--3   & 1.2$\pm$0.2 & 1.8$\pm$0.4 & 0.9$\pm$0.1 \\
Cha~I                          &  2--3   & 1.0$\pm$0.1 & 1.8$\pm$0.3 & 0.8$\pm$0.1 \\
$\sigma$~Orionis        &  3--5   & 1.0$\pm$0.2 & 2.0$\pm$0.4 & 0.6$\pm$0.1 \\
Upper Sco                   &  5--11 & 0.8$\pm$0.2 & 2.4$\pm$0.4 & 0.7$\pm$0.1
\enddata 
\tablenotetext{}{$^{\dagger}$Fit taken from \cite{2016ApJ...828...46A}, as they used the same methodology described in Section~\ref{sec-slopes}, but also an MC analysis to account for 20 Lupus sources with unknown stellar masses.} 
\tablenotetext{}{$^{\ddagger}$We use the convention of \cite{2007ApJ...665.1489K}, where $\beta$ and $\alpha$ represent the slope and intercept, respectively. This differs from that of \cite{2016ApJ...831..125P}, who switched these symbols.}
\end{deluxetable} 
\capstartfalse

To compare our $M_{\rm dust}$--$M_{\ast}$ relation derived for $\sigma$~Orionis to those found for other star-forming regions in a consistent manner, we follow the procedure described in \cite{2016ApJ...828...46A}. Namely, we calculate $M_{\rm dust}$ uniformly across each region by inputting the (sub-)mm continuum fluxes (or $3\sigma$ upper limits) from the literature into Equation~\ref{eqn-mass}, along with the cluster distances and observation wavelengths of the surveys. We assume $T_{\rm dust}=20$~K for all disks and adopt distances of 140~pc for Taurus \citep{2008hsf1.book..405K}, 150~pc or 200~pc for Lupus \citep{2008hsf2.book..295C}, 160~pc for Chamaeleon~I \citep{2008hsf2.book..169L}, and 145~pc for Upper Sco \citep{1999AJ....117..354D}. For Upper Sco, we only include the ``full," ``evolved," and ``transitional" disks from the sample of \cite{2016ApJ...827..142B}, as these represent the ``primordial" disks that do not yet show signs of disk clearing. Only sources with $M_{\ast}\ge0.1~M_{\odot}$ were considered in order to exclude brown dwarfs, while also maintaining a common stellar mass limit among the surveys. Stellar masses taken from the literature were derived using the \cite{2000A&A...358..593S} evolutionary tracks for all regions except Chamaeleon~I, which used the \cite{2015A&A...577A..42B} models; the stellar masses derived using these two grids are generally consistent, thus any effects should be negligible.

The fitted linear regression parameters for each region are given in Table~\ref{tab-linmix} and plotted in Figure~\ref{fig-mdust}. Our results further support the steepening of the $M_{\rm dust}$--$M_{\ast}$ relation with age, although the errors are large. Interestingly, we also find similarly large intrinsic dispersions for all five regions; as previously noted by \cite{2016ApJ...831..125P}, this seems to be an inherent property of disk populations, reflecting a range of disk conditions (e.g., dust opacities, disk evolutionary states, dust temperatures) rather than the age and/or environment of the region, and may partially account for the diversity seen in the exoplanet population. 

We note that our fitted values in Table~\ref{tab-linmix} are mostly consistent with those found by \cite{2016ApJ...831..125P} (see their Table~4) despite differences in assumptions of grain opacity (e.g., they use $\beta=0.4$ in Equation~\ref{eqn-mass} while we use $\beta=1.0$) and stellar mass cutoffs (e.g., they include sources with $M_{\ast}<0.1~M_{\odot}$, while we exclude brown dwarfs). Indeed, the main disagreement is the intercept estimate for Lupus, which differs because \cite{2016ApJ...831..125P} exclude the 20 sources in Lupus with unknown stellar masses, while we account for them using the MC approach described in \cite{2016ApJ...828...46A}. The slope for Upper Sco is also noticeably different (although within errors) because \cite{2016ApJ...831..125P} only consider ``full" and ``transitional" disks from Upper Sco, while we also include ``evolved" disks following the definition of ``primordial" disks in \cite{2016ApJ...827..142B}.

Finally, we address three potential caveats to our Bayesian linear regression fit to $\sigma$~Orionis disks. First, \cite{2016ApJ...831..125P} found that shallower slopes can result when the sample is dominated by upper limits at low stellar masses. Although roughly two-thirds of our $\sigma$~Orionis sample was undetected in the continuum, the non-detections span a range of stellar masses (see Figure~\ref{fig-mdust}); moreover, even if the slope is actually steeper, this would only further distinguish $\sigma$~Orionis from the younger regions. Second, the source with the highest stellar mass in $\sigma$~Orionis is not detected in the continuum, which could potentially skew the fit. We could not find a valid reason for discarding this source (e.g., no evidence for binarity), although its spectral type is uncertain \cite[$\pm$2.5 spectral types;][]{2014ApJ...794...36H}. Nevertheless, when re-fitting the distribution without this source, we recover the same parameters. Third, we have shown in Section~\ref{sec-photoevap} that external photoevaporation is reducing disk dust masses in $\sigma$~Orionis. This may serve to steepen the $M_{\rm dust}$--$M_{\ast}$ relation, as external photoevaporation should be more effective at removing gas and small dust grains around lower-mass stars whose orbiting material is less gravitationally bound. Unfortunately this possible effect has not yet been tested with external photoevaporation models. Given these potential issues, a more sensitive (sub-)mm continuum survey of $\sigma$~Orionis, as well as additional observations and detailed modeling, will be needed to confirm the linear regression fit presented here. Furthermore, we note that several other linear regression methods for left-censored datasets are available in the statistical literature and should be tested for this specific astronomical problem.

\subsection{Dust Mass Distributions \label{sec-dustdist}}

\capstartfalse
\begin{figure}
\begin{centering}
\includegraphics[width=8.5cm]{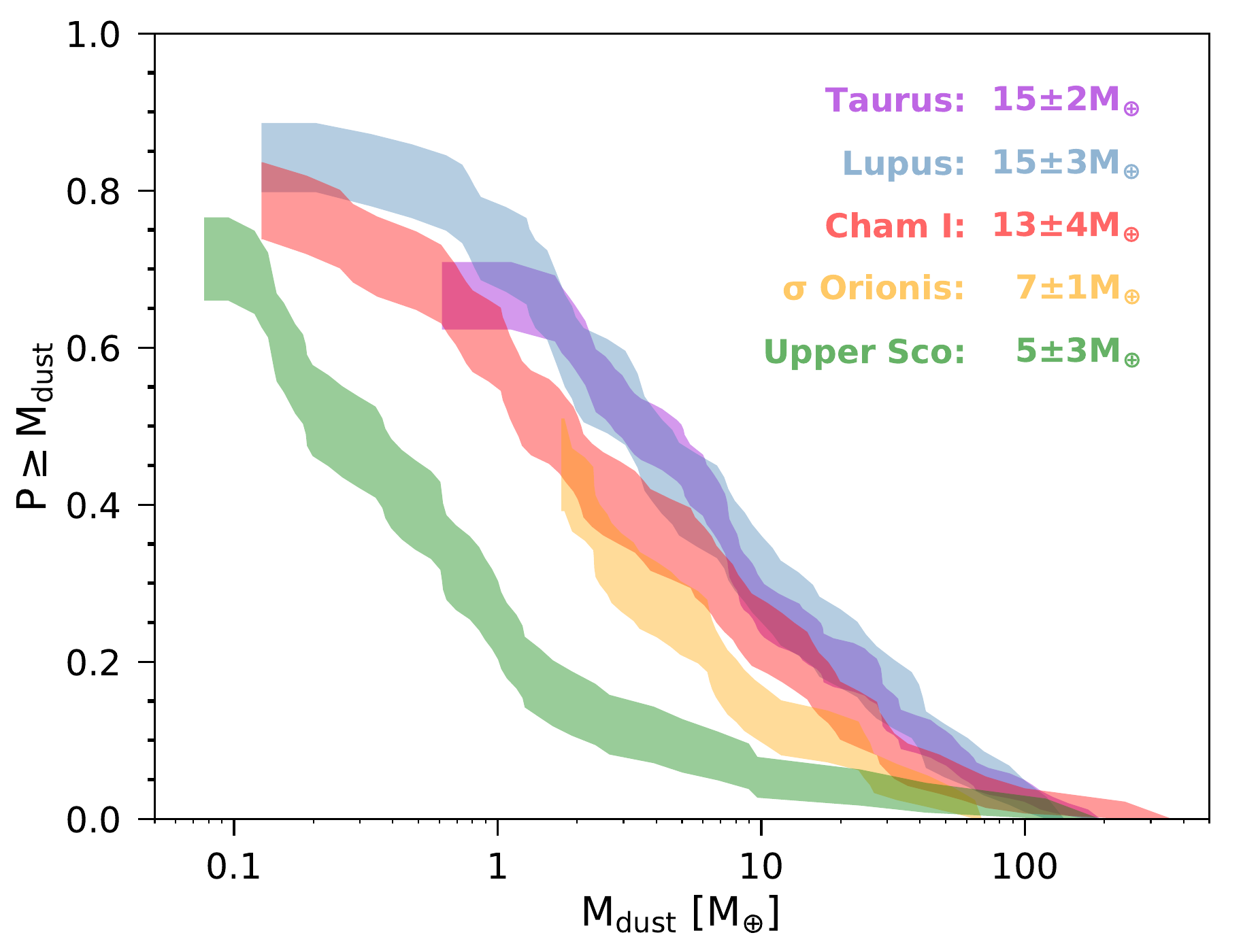}
\caption{\small Dust mass cumulative distributions for Taurus, Lupus, Chamaeleon~I, $\sigma$~Orionis, and Upper Sco disks around host stars with $M_{\ast}\ge0.1~M_{\odot}$. The average dust mass for each region is given for reference. The distributions and their 1$\sigma$ confidence intervals were calculated using the Kaplan-Meier estimator in the \texttt{ASURV} package \citep{1992ASPC...25..245L} to properly account for upper limits using well-established techniques for left-censored datasets.}
\label{fig-dustdist}
\end{centering}
\end{figure}
\capstartfalse

Disk dispersal and grain growth should be reflected in a steady decline with age of the bulk dust mass probed by (sub-)mm continuum flux. \cite{2016ApJ...828...46A} showed that the average dust mass of disks in a given star-forming region, as measured from (sub-)mm continuum flux, does indeed decline with age. Specifically, they found that Lupus and Taurus have consistent mean dust masses ($15\pm3$~$M_{\oplus}$ and $15\pm2$~$M_{\oplus}$, respectively), while Upper Sco has a mean dust mass that is $\sim$5$\times$ lower ($5\pm3$~$M_{\oplus}$). They calculated dust masses uniformly across each region (as described in Section~\ref{sec-slopes}) then derived the average dust mass using the Kaplan-Meier estimator in the \texttt{ASURV} package \citep{1992ASPC...25..245L} to properly account for the upper limits using well-established techniques for left-censored datasets. We follow this procedure to now include results from our ALMA survey of $\sigma$~Orionis disks as well as the ALMA survey of Chamaeleon~I disks \citep{2016ApJ...831..125P}. As shown in Figure~\ref{fig-dustdist}, we find average disk dust masses of $7\pm1$~$M_{\oplus}$ for $\sigma$~Orionis and $13\pm4$~$M_{\oplus}$ for Chamaeleon~I, further supporting the decline of this parameter with age. 
 
When comparing the dust mass distributions of two regions, it is important to confirm that they have similar stellar mass distributions due to the $M_{\rm dust}$--$M_{\ast}$ relation discussed in Section~\ref{sec-slopes}. We therefore employ the two-sample tests in the \texttt{ASURV} package to determine the probabilities that the stellar masses of each region are drawn from the same parent population. We find that the stellar mass distributions are statistically indistinguishable for $\sigma$~Orionis and Lupus ($p=0.45$--0.96), $\sigma$~Orionis and Chamaeleon~I ($p=0.06$--0.22), and $\sigma$~Orionis and Upper Sco ($p=0.21$--0.30). However we found statistically distinct stellar mass distributions for $\sigma$~Orionis and Taurus ($p=0.03$--0.04). Therefore, we can directly compare the dust mass distribution of $\sigma$ Orionis to those of Lupus,  Chamaeleon~I, and Upper Sco in Figure~\ref{fig-dustdist}. However, caution should be taken when interpreting the comparison of $\sigma$~Orionis to Taurus in Figure~\ref{fig-dustdist} due to their potentially different stellar mass distributions. Nevertheless, our main conclusions in the previous paragraph remain the same.

Alternatively, to account for stellar mass differences between regions, one could instead compare the $M_{\rm dust}/M_{\ast}$ distributions \cite[e.g.,][]{2016ApJ...827..142B} or employ an MC approach that aims to normalize the stellar mass selection functions \cite[e.g.,][]{2013ApJ...771..129A,2013MNRAS.435.1671W,2016ApJ...828...46A}. However, in this work we do not attempt these more detailed analyses given the larger uncertainties on our $M_{\ast}$ estimates, especially for the sources with photometrically derived stellar masses.

\subsection{Ingredients for Giant Planet Formation \label{sec-planets}}

Core accretion theory predicts that giant planets form when solid cores of a minimum critical mass assemble in the disk, enabling runaway accretion of the surrounding gaseous material  \citep{1996Icar..124...62P,2004ApJ...604..388I}. The accretion of a gaseous envelope is expected to occur rapidly, where $\sim$10~$M_{\oplus}$ cores reach masses of $\sim$1~$M_{\rm Jup}$ within $\sim$0.1~Myr. Within the framework of this model, we can constrain the occurrence of giant planet formation by observing how quickly the dust and gas content in typical protoplanetary disks depletes to levels below what are thought to be needed to form a gas giant.

In particular, we can look at the fraction of protoplanetary disks in a region with dust masses above the $\sim$$10~M_{\oplus}$ limit needed to form a giant planet core. For regions at $\sim$1--3~Myr, we see roughly a quarter of protoplanetary disks above this threshold (30\% in Taurus, 26\% in Lupus, and 23\% in Chamaeleon~I). At $\sim$3-5~Myr, we see this fraction cut in half (13\% in $\sigma$~Orionis) and then halved again at $\sim$5--10~Myr (5\% in Upper Sco). Although these are only rough estimates due to different survey completenesses, they seem to clearly reflect a sharp decline in the capacity of disks to form giant planets with age. Even in the youngest regions, the majority of disks appear to lack sufficient dust to form the solid cores needed to build giant planets, implying that giant planet formation is either rare or well on its way after just a few Myr.

Additionally, stacking (sub-)mm continuum non-detections allows us to put limits on the average amount of dust in the lowest-mass disks (e.g., Section~\ref{sec-stacks}). In Lupus, \cite{2016ApJ...828...46A} found that the undetected disks had extremely low average dust masses of $\lesssim6$ Lunar masses (0.03~$M_{\oplus}$), comparable to debris disk levels \citep{2008ARA&A..46..339W}. Although the further distance of $\sigma$~Orionis results in looser constraints, we still find that undetected disks have $\lesssim4$ Martian masses (0.4~$M_{\oplus}$) of dust on average (Section~\ref{sec-stacks}). These findings support theoretical predictions that viscous disks evolve rapidly into debris disks once stellar accretion ceases and photoevaporation from the central star dominates, clearing the dust from the inside out and leaving behind larger solids such as pebbles and planetesimals \cite[e.g.,][]{2001MNRAS.328..485C, 2006MNRAS.369..229A}. Previous studies of weak-lined T Tauri stars have provided observational evidence for rapid disk clearing \citep[e.g.,][]{2013ApJ...762..100C,2015A&A...583A..66H}, and \cite{2013MNRAS.435.1037P} used a compilation of sub-mm fluxes from the literature to show that debris disks are substantially less massive than disks around younger pre-main sequence stars. However, our larger and more homogeneous samples of Lupus and $\sigma$~Orionis disks confirm that rapid disk clearing is a uniform occurrence, even among young protoplanetary disk populations. 

Another ingredient for giant planet formation is of course the gas. Although bulk gas masses are notoriously difficult to measure (Section~\ref{sec-gas}), the fact that we found only three disks in $\sigma$~Orionis that exhibit both $^{12}$CO and $^{13}$CO emission is telling; indeed, their line fluxes correspond to gas masses of just $\sim$2--7~$M_{\rm Jup}$ using the methodology described in Section~\ref{sec-gas}. For the remaining disks in $\sigma$~Orionis, we find upper limits on their individual gas masses of just $\sim$3~$M_{\rm Jup}$. Moreover, the average gas mass of disks detected in the continuum but undetected in $^{12}$CO is $<1.0~M_{\rm Jup}$ (Section~\ref{sec-stacks}). These low gas masses again suggest that giant planet formation is either rare or nearly complete by the $\sim$3--5~Myr age of $\sigma$ Orionis (or that carbon is being significantly depleted in protoplanetary disks; see Section~\ref{sec-gas}).

\capstartfalse
\begin{figure}
\begin{centering}
\includegraphics[width=8.8cm]{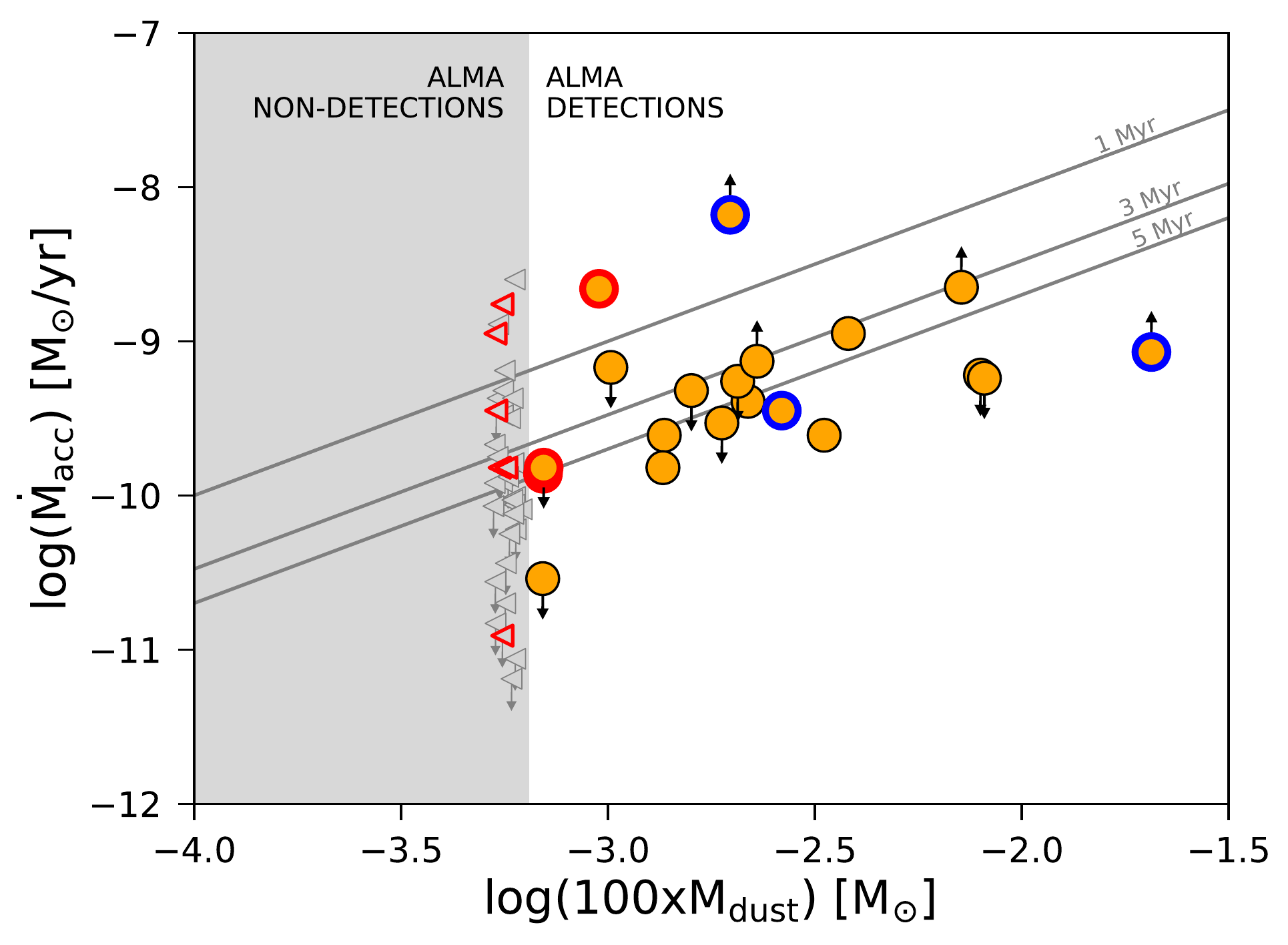}
\caption{\small Stellar mass accretion rate ($\dot{M}_{\rm acc}$) from the $U$--band survey of \cite{2011A&A...525A..47R} versus disk mass ($M_{\rm disk} = 100 \times M_{\rm dust}$) from our ALMA observations, for the $\sigma$~Orionis members included in both surveys. Orange circles are ALMA continuum detections and gray triangles are 3$\sigma$ upper limits. Upward/downward arrows are lower/upper limits on $\dot{M}_{\rm acc}$. Sources outlined in blue are our CO detections (Section~\ref{sec-line}) and sources outlined in red are $\lesssim0.5$~pc from the central OB system (Section~\ref{sec-photoevap}). The diagonal lines show different $M_{\rm disk}$/$\dot{M}_{\rm acc}$ ratios, which represents different viscous timescales.}
\label{fig-macc}
\end{centering}
\end{figure}
\capstartfalse

\subsection{Relation between $M_{\rm disk}$ and  $\dot{M}_{\rm acc}$ \label{sec-viscous}}

Viscously evolving protoplanetary disks should exhibit a direct relation between their total disk mass ($M_{\rm disk}$) and stellar mass accretion rate ($\dot{M}_{\rm acc}$), as shown in \cite{1998ApJ...495..385H}. This theoretical prediction was only recently confirmed observationally by \cite{2016A&A...591L...3M}, who combined the ALMA \citep{2016ApJ...828...46A} and VLT/X-Shooter \citep{2014A&A...561A...2A,2016arXiv161207054A} surveys of protoplanetary disks in the Lupus clouds to reveal a linear relation between $\dot{M}_{\rm acc}$ and $M_{\rm disk}$. \cite{2016A&A...591L...3M} also showed that, when assuming an ISM gas-to-dust ratio of $\sim$100, the $M_{\rm disk}$/$\dot{M}_{\rm acc}$ ratio is consistent with the age of the region, as expected for viscously evolving disks \citep{2017arXiv170302974R}. These observational findings suggest that mass accretion onto the stellar surface is indeed related to the properties of the outer disk. 

To search for a similar correlation in $\sigma$~Orionis, we combine our ALMA observations with the $U$--band survey of \cite{2011A&A...525A..47R}. The latter used their photometric data to estimate $\dot{M}_{\rm acc}$, making these accretion rates more uncertain than those obtained with spectroscopy for Lupus disks. As shown in Figure~\ref{fig-macc}, only 20 sources (the most massive disks in $\sigma$~Orionis) have both $M_{\rm dust}$ estimates and constraints on $\dot{M}_{\rm acc}$ (the latter of which are mostly upper or lower limits). The sparseness of this sub-sample, combined with the smaller $M_{\rm dust}$ range when compared to the Lupus sample, limits our ability to fit a relation similar to that in \cite{2016A&A...591L...3M}. However, we note that, as for Lupus disks, the $M_{\rm disk}$/$\dot{M}_{\rm acc}$ ratios are consistent with the age of $\sigma$~Orionis for an ISM gas-to-dust ratio of $\sim$100.

The disks undetected with ALMA (i.e., the least massive disks in $\sigma$~Orionis) span over 2 dex in $\dot{M}_{\rm acc}$, and thus a correspondingly large range of $M_{\rm disk}$/$\dot{M}_{\rm acc}$ ratios, as shown in Figure~\ref{fig-macc}. The undetected disks with low $\dot{M}_{\rm acc}$ values or upper limits are consistent with expectations from viscous evolution. However, the undetected disks with moderate-to-high $\dot{M}_{\rm acc}$ values ($\gtrsim 2\times 10^{-10} M_\odot$~yr$^{-1}$) are unexpected, as these should have lifetimes shorter than the age of the region. The sources $\lesssim0.5$~pc from the central OB star (outlined in red in Figure~\ref{fig-macc}) are most readily explained by external photoevaporation (see Section~\ref{sec-photoevap}), which would serve to reduce $M_{\rm disk}$ by removing mass from the outer disk, thereby increasing the $M_{\rm disk}/\dot{M}_{\rm acc}$ ratio \citep{2017arXiv170302974R}. Although this accounts for only a handful of objects, there is evidence that external photoevaporation is occurring throughout the region (Section~\ref{sec-photoevap}), thus may apply to more of the undetected disks. Additionally, the $\dot{M}_{\rm acc}$ values estimated from $U$-band photometry are uncertain and need to be confirmed with spectroscopy. Nevertheless, there are a sufficient number of sources with low disk masses and significant accretion rates to warrant further investigation. These objects might have strongly variable accretion rates, or alternatively the accreting gas may come from the evaporation of ice-coated dust grains.


\section{Summary}
\label{sec-summary}
 
We have used ALMA to conduct a high-sensitivity mm survey of protoplanetary disks in the $\sigma$~Orionis cluster. This region is particularly interesting for studying disk evolution as its intermediate age ($\sim$$3$--5~Myr) is comparable to the median disk lifetime, and therefore corresponds to a potentially important phase of disk evolution and planet formation. 

\begin{enumerate}

\item We used ALMA to survey the dust and gas in 92 protoplanetary disks around $\sigma$~Orionis members with $M_{\ast}\gtrsim0.1~M_{\odot}$. Our observations cover the 1.33~mm continuum as well as the $^{12}$CO, $^{13}$CO, and C$^{18}$O $J=2$--1 lines. Out of the 92 sources, we detected only 37 in the mm continuum and six in $^{12}$CO, three in $^{13}$CO, and none in C$^{18}$O. 

\item The continuum emission constrained dust masses to $\sim$2~$M_{\oplus}$, while the CO line emission constrained gas masses to $\sim$3~$M_{\rm Jup}$. Only 11 disks had $M_{\rm dust}\gtrsim10~M_{\oplus}$, indicating that after a few Myr of evolution the vast majority of disks lack sufficient dust to form giant planet cores. The low gas masses also indicate that giant planet formation must be rapid or rare, but may also reflect significant carbon depletion in protoplanetary disks. Moreover, stacking the individually undetected continuum sources limited their average dust mass to $\sim$5$\times$ lower than that of the faintest detected disk, supporting theoretical models that predict disks dissipating rapidly once accretion stops and photoevaporation dominates. 

\item We found that external photoevaporation from the central OB stars is influencing disk evolution throughout the region. Namely, disk dust masses clearly decline with decreasing projected separation from the photoionizing source, and the handful of cluster members with detected CO emission exist only at projected separations $>1.5$~pc. This indicates that even moderate external FUV fluxes can result in significant mass-loss rates, and future theoretical studies will be needed to quantify the implications for planet formation in OB clusters.

\item Comparing the protoplanetary disk population in $\sigma$~Orionis to those of other star-forming regions provided continuing support of the steady decline in average disk dust mass and steepening of the $M_{\rm dust}$--$M_{\ast}$ relation with age. Quantifying these evolutionary trends can help to determine the relative importance of different disk processes during key eras of planet formation. However, for $\sigma$~Orionis, these trends may also be influenced by the effects of external photoevaporation from the central OB stars.

\item Collectively, our findings indicate that giant planet formation is inherently rare and/or well underway by a few Myr of age. However, due to the abundance of upper limits in our ALMA sample, and the need for better constraints on stellar properties, a higher-sensitivity (sub-)mm survey as well as a complete spectroscopic survey of the members of $\sigma$~Orionis analyzed in this work should be conducted to confirm these results.

\end{enumerate}

 
\begin{acknowledgements}

MCA and JPW were supported by NSF and NASA grants AST-1208911 and NNX15AC92G, respectively. MCA acknowledges student observing support from NRAO.  CFM acknowledges an ESA Research Fellowship. NM is supported in part by the Beatrice W. Parrent Fellowship in Astronomy at the University of Hawaii. Leiden is supported by the European Union A-ERC grant 291141 CHEMPLAN, by the Netherlands Research School for Astronomy (NOVA), and by grant 614.001.352 from the Netherlands Organization for Scientific Research (NWO). This paper makes use of the following ALMA data: ADS/JAO.ALMA2015.1.00089.S. ALMA is a partnership of ESO (representing its member states), NSF (USA) and NINS (Japan), together with NRC (Canada), NSC and ASIAA (Taiwan), and KASI (Re- public of Korea), in cooperation with the Republic of Chile. The Joint ALMA Observatory is operated by ESO, AUI/NRAO and NAOJ. The National Radio Astronomy Observatory is a facility of the National Science Foundation operated under cooperative agreement by Associated Universities, Inc.

\end{acknowledgements}


\bibliography{../../bib.bib}

\end{document}